\documentclass[twocolumn, superscriptaddress, 
floatfix]{revtex4-2}

\usepackage{graphicx} 
\usepackage{float}
\usepackage{mathrsfs}
\usepackage{siunitx}
\usepackage{subcaption}
\usepackage{subfiles} 
\usepackage{dcolumn} 
\usepackage{bibunits}
\defaultbibliography{references}
\defaultbibliographystyle{apsrev4-2}
\usepackage[x11names]{xcolor}
\usepackage[x11names]{xcolor}
\definecolor{darkgreen}{rgb}{0.0, 0.5, 0.0}
\definecolor{yellow}{rgb}{0.9, 0.5, 0}
\definecolor{darkred}{rgb}{0.8, 0, 0}
\definecolor{darkblue}{rgb}{0, 0, 0.6}
\usepackage{chngcntr}
\usepackage{amsmath}
\usepackage{placeins}
\usepackage{ulem}
\usepackage{setspace}
\usepackage{comment}
\usepackage{ragged2e}
\usepackage{lineno}
\usepackage{xspace}
\usepackage{hyperref}
\begin{document}
\begin{bibunit}

\newcommand{\MoTe}{MoTe$_2$\xspace}
\newcommand{\mvnm}{\milli \volt \per \nano \meter}
\newcommand{\RGhBN}{R$5$G/hBN\xspace}

\title{$1/3$ Fractional and Gapless Integer Quantum Anomalous Hall States in \\
Rhombohedral Graphene}
\author{Jackson P. Butler}
\altaffiliation{These authors contributed equally}
\affiliation{Department of Physics, Massachusetts Institute of Technology, Cambridge, MA, 02139, USA}
\author{Tonghang Han}
\altaffiliation{These authors contributed equally}
\affiliation{Department of Physics, Massachusetts Institute of Technology, Cambridge, MA, 02139, USA}
\author{Andrew DiFabbio}
\affiliation{Department of Physics, Massachusetts Institute of Technology, Cambridge, MA, 02139, USA}
\author{Zach Hadjri}
\affiliation{Department of Physics, Massachusetts Institute of Technology, Cambridge, MA, 02139, USA}
\author{Emily Aitken}
\affiliation{Department of Physics, Massachusetts Institute of Technology, Cambridge, MA, 02139, USA}
\author{Kenji Watanabe}
\affiliation{Research Center for Electronic and Optical Materials, National Institute for Materials Science, 1-1 Namiki, Tsukuba 305-0044, Japan}
\author{Takashi Taniguchi}
\affiliation{Research Center for Materials Nanoarchitectonics, National Institute for Materials Science, 1-1 Namiki, Tsukuba 305-0044, Japan}
\author{Long Ju}
\altaffiliation{longju@mit.edu}
\affiliation{Department of Physics, Massachusetts Institute of Technology, Cambridge, MA, 02139, USA}
\author{Raymond C. Ashoori}
\altaffiliation{rashoori@mit.edu}
\affiliation{Department of Physics, Massachusetts Institute of Technology, Cambridge, MA, 02139, USA}

\begin{abstract}

The fractional quantum anomalous Hall (FQAH) effect occurs in moiré superlattices in both twisted bilayer \MoTe (t\MoTe)\cite{park_observation_2023, xu_observation_2023} and rhombohedral $n$-layer graphene aligned to hexagonal boron nitride (R$n$G/hBN)\cite{lu_fractional_2024, lu_extended_2025, xie_tunable_2025} as a novel quantum phase driven by intertwined electron correlation and topology. Although several fractional states in the Jain sequence\cite{jain_compositefermion_1989} have been identified, the $1/3$ state, the most robust and fundamental state in conventional fractional quantum Hall (FQH) systems\cite{laughlin_anomalous_1983}, was missing in either FQAH system. Determining whether or not it exists would have a major impact on understanding the mechanism of FQAH and its relation to the conventional Landau level picture, especially in the theoretically still-debated R$n$G/hBN system.  Here we report the FQAH effect at moiré filling factor $\nu = 1/3$ in \RGhBN moiré superlattice devices, through a combination of quantum capacitance and transport measurements. By tuning the displacement field, we observed a topological phase transition from a $1/3$ fractional Chern insulator (FCI) to a trivial charge density wave state. We establish thermodynamic gaps of several FCI states including the largest of $\Delta\mu \sim \SI{7}{\kelvin}$ at $\nu = 1/3$. With the inclusion of the $1/3$ state, the FQAH states in \RGhBN now exhibit a surprising level of particle-hole symmetry about half-filling of the moiré band, closely resembling the behavior of FQH states in the lowest Landau level. Additionally, we perform compressibility and transport measurements at a filling of one electron per moiré unit cell, $\nu =1$, and also for $\nu \lesssim 1$, where previous transport measurements displayed the extended quantum anomalous Hall (EQAH) effect\cite{lu_extended_2025}. While our transport measurements show no change between the integer quantum anomalous Hall state (IQAH) and the EQAH region, compressibility measurements reveal a distinct transition from a gapped IQAH state to a gapless and highly compressible EQAH state. Our direct thermodynamic characterization of the rich phase diagram paves the way to engineering of anyon braiding and non-Abelian quasiparticles at zero magnetic field.





\end{abstract}
\maketitle
\begin{figure*}[t]
  \includegraphics[width=\linewidth]{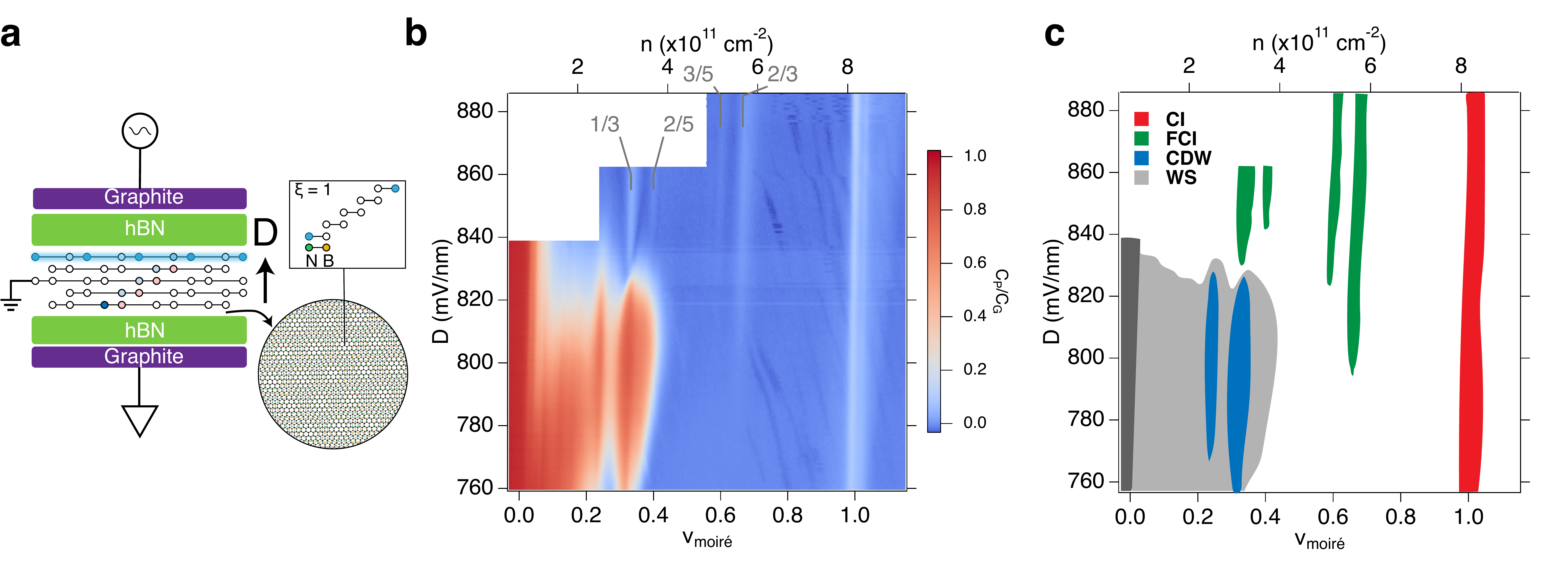}
  \caption{
      \justifying \textbf{Phase diagram of R$5$G/hBN system in the moiré-distant regime} \textbf{a)} Diagram of the device and measurement schematic. An AC excitation is applied to the top gate and the penetrating signal is measured on the back gate. A moiré superlattice is formed by aligning the graphene to the bottom hBN substrate due to the small lattice mismatch between the two materials. The stacking orientation of the hBN is such that the $A$ site carbon atom sits on top of the nitrogen atom and is labeled as $\xi =1$ as defined by ref. \cite{uzan_hbn_2025}. The arrow depicts the direction of the applied DC displacement field that pushes electrons to the opposite side of the R$5$G from the moiré pattern. \textbf{b)} Phase diagram of the device as measured in penetration capacitance $C_P$ normalized by the geometric capacitance, $C_G$. At large displacement field and charge neutrality, a single particle gap opens and $C_P = C_G$. The frequency of the applied excitation is $\SI{10}{\mega \hertz}$ and the temperature of this refrigerator is $\SI{320}{\milli \kelvin}$. There are extraneous background diagonal features in the upper right portion of the figure arising from a bad contact. These are diminished by increasing the measurement frequency. (See Supplementary Information \ref{sec:SI_frequencyDependence} for more information). The upper left portions of the map are blank due to gate leakage in this regime. \textbf{c)} Cartoon phase diagram outlining the different states seen in the penetration capacitance measurement.
    }
  \label{PhaseDiagram}
\end{figure*}

Recently, the fractional quantum anomalous Hall effect (FQAH) has been observed in both twisted bilayer \MoTe (t\MoTe)\cite{park_observation_2023, xu_signatures_2025, xu_observation_2023}, and rhombohedral $n$-layer graphene aligned to a hexagonal boron nitride (hBN) substrate for $n = 4-6$ (R$n$G/hBN)\cite{lu_fractional_2024, lu_extended_2025, xie_tunable_2025}. The underlying mechanism that gives rise to the effect in these two systems appears to be fundamentally different. t\MoTe features both spin-orbit coupling and strong moiré potential which give rise to an isolated Chern band\cite{wu_topological_2019, li_spontaneous_2021, yu_giant_2020, devakul_magic_2021}. In contrast, the FQAH states in \RGhBN only appear when the electrons are pushed away from the graphene/hBN moiré interface, the moiré-distant regime. Further, recent theory suggests that interactions are necessary to create an isolated Chern band in the moiré-distant limit\cite{tan_ideal_2025, li_multiband_2025, yu_moire_2025, huang_fractional_2025, zhou_fractional_2024, dong_theory_2024, huang_selfconsistent_2024, guo_fractional_2024, xie_integer_2024, song_intertwined_2024}. This leaves open the question about how the underlying mechanism of FQAH in \RGhBN is related to the fractional quantum Hall (FQH) effect observed in the lowest Landau level (LLL) at high magnetic field. The FQAH states previously measured by transport experiments all belong to the Jain sequence, which describes the high-field conventional FQH effect\cite{jain_compositefermion_1989}. However, the $1/3$ FQAH state, which is the most fundamental state in the conventional FQH effect and is described by the Laughlin wavefunction, has not previously been observed at zero magnetic field in any system.  Further, both FQAH systems, t\MoTe and \RGhBN, exhibit either extended (EQAH) or re-entrant  integer quantum anomalous Hall (RIQAH) effects which display quantized Hall resistances for ranges of filling factors outside of $\nu =1$ \cite{lu_extended_2025, xu_signatures_2025}. However, very little is known about the bulk electronic properties of either of these phases.


The lack of experimental studies outside of conventional transport has left unanswered many questions about the nature of the FQAH and EQAH. Unlike transport, penetration capacitance is a measurement of the bulk electronic properties of a 2d system. It probes the thermodynamic density of states of the bulk and thus can identify incompressible phases and quantify the size of the energy gap. Understanding the hierarchy of the energy gaps of the fractional states would elucidate the relationship of the FQAH effect to the conventional FQH effect. Further, these measurements provide valuable information about the compressibility of the EQAH and RIQAH phases, as well as allowing detection of highly resistive states where transport fails. 

Here we investigate \RGhBN in the moiré-distant limit with both penetration capacitance and transport techniques. We observed a robust FQAH state at $1/3$ filling, which was not previously measured in either t\MoTe or \RGhBN system. The state undergoes a topological phase transition from a FQAH state to a trivial charge density wave (CDW) state as the displacement field is decreased. Moreover, the $1/3$ state has the largest thermodynamic gap of any of the fractional states in the system. The phase diagram as measured by compressibility reveals that the fractional states display substantial particle-hole symmetry about half-filling of the moiré band. Further, at the lowest temperatures, we measure a region of near-zero longitudinal resistance extending from $\nu = 1$ to lower densities, corresponding to the previously measured EQAH effect. The compressibility shows a crossover from an incompressible state at $\nu = 1$, as expected for an integer quantum anomalous Hall (IQAH) state, to strong negative compressibility for fillings below $\nu = 1$ thus showing that the EQAH regime lacks a bulk thermodynamic gap.

\begin{figure*}[t]
  \includegraphics[width=\linewidth]{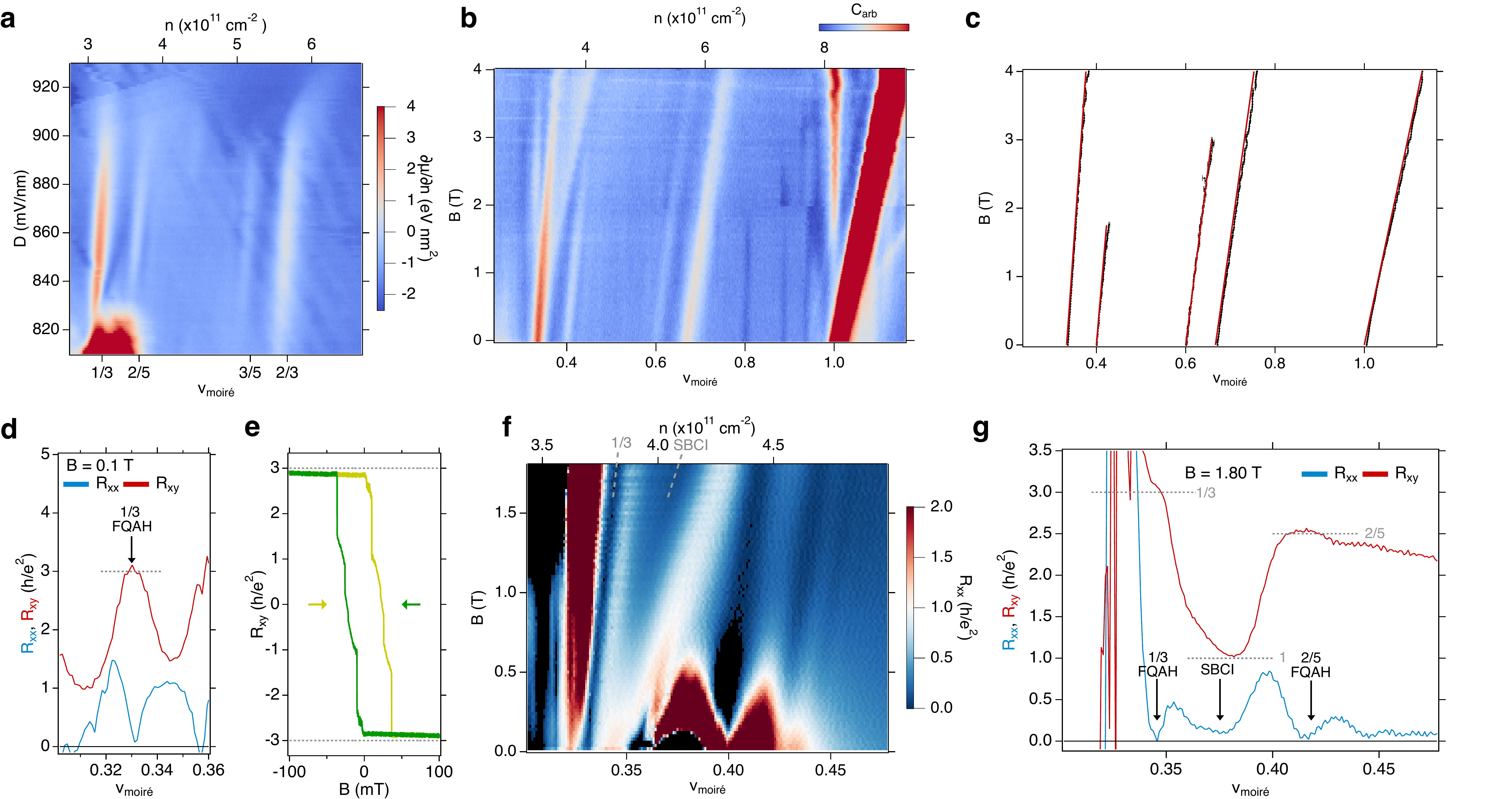}
  \caption{%
      \justifying
      {\textbf{Compressibility and Transport of the Fractional Quantum Anomalous Hall states} \textbf{a)} Compressibility extracted from the capacitance measurement of the fractional Chern insulator states at $T = \SI{39}{\milli \kelvin}$ and $B = \SI{0}{\tesla}$. Capacitance measurement was taken at $f = \SI{202}{\kilo \hertz}$.  \textbf{b)} Landau fan measurement taken at $D = \SI{852}{\milli \volt \per \nano \meter}$ at $T = \SI{320}{\milli \kelvin}$ at $f = \SI{10}{\mega \hertz}.$  \textbf{c)}  Wannier diagram tracking the densities for each of the IQAH and FQAH states at each magnetic field. The red lines correspond to the expected slope of fractional Chern number states. \textbf{d)} $R_{xx}$ and $R_{xy}$ measurement as a function of density at $B = \SI{0.1}{\tesla}$ in device D2 showing the FQAH state. \textbf{e)} Magnetic hysteresis scan measuring $R_{xy}$ at the $\nu =1/3$ state in device D2. \textbf{f)} Landau fan measuring the longitudinal resistance, $R_{xx}$ in a second device D2. \textbf{g)} $R_{xx}$ and $R_{xy}$ measurement as a function of density at $B = \SI{1.8}{\tesla}$ in device D2. All transport data are symmetrized/antisymmetrized and measured at $T = \SI{300}{\milli \kelvin}$.}}
    \label{FCI_LandauFan}
\end{figure*}

\section{Phase Diagram}
Figure \ref{PhaseDiagram}\textbf{b} shows the penetration capacitance as a function of density and displacement field in the moiré-distant limit. At small filling factor, $\nu \lesssim 1/2$ and lower displacement field we observe a large penetration capacitance over an extended region of density and displacement field. We attribute this region of the map to a Wigner solid (WS) state, an electronic solid with short range order and high resistivity, as has previously been observed in transport measurements\cite{lu_fractional_2024, lu_extended_2025}. In the low-frequency limit, a long-ranged ordered Wigner crystal (WC) or a WS is thought to give rise to negative compressibility, or over-screening, due to strong electronic interactions\cite{eisenstein_negative_1992, eisenstein_compressibility_1994}. However, due to high in-plane resistivity, the system cannot fully charge for AC excitation periods faster than the charging time. In this regime, the penetration capacitance is no longer strictly a probe of the density of states, but rather a frequency dependent measurement of the compressibility and the in-plane resistivity\cite{goodall_capacitance_1985, dultz_thermodynamic_2000, zibrov_tunable_2017}. 


Within the WS region, we observe peaks in the penetration capacitance at commensurate filling factors such as $\nu = 1/3 \textrm{ and } 1/4$ as well as higher orders, which we attribute to trivial CDW states. Further, there is a weakly incompressible state with a dissapative component at $\nu =1/2$ (See Fig. \ref{Extended_Data_Closs_PhaseDiagram}) which may also be a CDW state. Unlike a WS, a CDW state is expected to have an associated thermodynamic gap. We confirm that these states are topologically trivial by examining the magnetic field dependence (see extended data Fig. \ref{nvsDvsB_maps}). Interestingly, in unaligned R$5$G, there is a region of very high resistance with a non-linear and hysteretic current-voltage relationship indicating a WC state around this same region of $n-D$ phase space\cite{han_evidence_2026}. Taken together with our measurements, this indicates that introducing a moiré pattern to the system helps pin the WC at commensurate filling factors to create CDW states. This suggests that the moiré pattern, even in the moiré-distant limit, is sufficiently strong at low electron densities to stabilize charge density wave states much like those seen in the moiré-proximal limit (see Fig. \ref{ExtendedData_MoireProximal} and ref. \cite{aronson_displacement_2025}).

As the displacement field is increased, the penetration signal of the $\nu = 1/3$ CDW state decreases before being replaced by a narrow incompressible state at $D \approx \SI{830}{\milli \volt \per \nano \meter}$ as shown in Fig. \ref{PhaseDiagram}\textbf{b}. This disappearance and reemergence in the capacitance suggests a phase transition between a trivial $1/3$ CDW state and a previously unobserved state at higher displacement field. Prior transport measurements in R$5$G/hBN revealed a highly resistive region extending out to nearly $\nu = 2/5$ which obscured $1/3$ filling\cite{lu_fractional_2024, lu_extended_2025}. As the density is increased we observe a sequence of incompressible states at commensurate filling factors $2/5$, $3/5$, and $2/3$ where FQAH states have previously been measured\cite{lu_fractional_2024, lu_extended_2025}. $C_P/C_G$ shows a sharp increase at $\nu = 1$, signaling a large thermodynamic gap where previous transport measurements have reported an IQAH state.

\section{Novel FQAH State}
To understand the topological nature of each of the incompressible states, we take a Landau fan measurement at $D = \SI{852}{\milli \volt \per \nano \meter}$, as shown in Fig. \ref{FCI_LandauFan}\textbf{b}. The Středa formula states that the density of a gapped state will change linearly in magnetic field according to the equation $\mathcal{C}  = \phi_0 \frac{\partial n}{\partial B}$, where $\phi_0$ is the magnetic flux quantum, and $\mathcal{C} $ is the Chern number. This can be related to the Hall conductivity through the relation $\sigma_{xy} = \mathcal{C} \frac{e^2}{h}$\cite{streda_quantised_1982, streda_theory_1982}. The newly identified incompressible state at $\nu =1/3$ has a slope which corresponds to $\mathcal{C}  = 0.329 \pm 0.001$, as shown in Fig. \ref{FCI_LandauFan}\textbf{c}, indicating a novel observation of a $1/3$ fractional Chern insulator. At larger filling, the slope of the incompressible states at fractional fillings of $\nu  = 2/5\textrm{, } 3/5\textrm{, and } 2/3$ correspond to $\mathcal{C}  = 0.465\pm 0.007 , 0.602\pm 0.008, \textrm{ and } 0.691 \pm 0.002$, indicating that these incompressible states correspond to previously identified FQAH states (See supplementary information for more detail). The extracted Chern numbers agree well with the previously reported anomalous Hall resistivities\cite{lu_fractional_2024, lu_extended_2025}. Thus, the incompressible states as shown in Fig. \ref{FCI_LandauFan}\textbf{a} correspond to fractional Chern insulator states. 

To further verify the topological nature of the $1/3$ state, we measure transport in a second device (D2) with a moiré period of \SI{11.5}{\nano \meter} that maintains good electrical contact down to low electron density. This is the same device from refs.\cite{lu_fractional_2024, lu_extended_2025}. We find that raising the temperature to $\SI{300}{\milli \kelvin}$ is necessary to observe the state in transport, due to the high resistance of the WS state surrounding $\nu = 1/3$. This potentially explains why the 1/3 FQAH state had not been seen before in RnG/hBN. Poor electrical contacts at low density prevent us from making transport measurements on device D1 below $\nu = 1/2$. Figure \ref{FCI_LandauFan}\textbf{d} and \textbf{e} show an anomalous Hall resistance in D2 that is quantized to nearly $R_{xy}  = 3 \frac{h}{e^2}$ with clear magnetic hysteresis. Further, Landau fan transport measurements shown in Fig. \ref{FCI_LandauFan}\textbf{f} reveal a vanishing longitudinal resistance along a slope that follows the Středa formula for $\mathcal{C} = 1/3$. These measurements in a second device, with a different moiré periodicity, reveal a dissipationless edge state with the  expected quantized Hall resistance, confirming that the thermodynamic gap at $\nu = 1/3$ corresponds to a FQAH state. 


Aside from fractional and integer quantum anomalous Hall states, we observe several other states that emerge at non-zero magnetic field. Trivial correlated insulator states at both $\nu = 1/3$ and $\nu = 1$ emerge at non-zero magnetic field, as displayed in Fig. \ref{FCI_LandauFan}\textbf{b} and \textbf{f}. These coexist with the corresponding FQAH or IQAH topological states. Strikingly, we also measure an incompressible state with a slope of $\mathcal{C} = 1$ emanating from $\nu = 1/3$. This $\mathcal{C} = 1$ state is also observed in device D2 with the expected slope and a quantized Hall resistance of approximately $h/e^2$ as shown in a linecut at $B = \SI{1.8}{\tesla}$ in  Fig. \ref{FCI_LandauFan}\textbf{g} (see extended data Fig. \ref{ExtendedData_RxyLandauFan}). We attribute this incompressible state to a $\mathcal{C}=1$ symmetry-broken Chern insulator (SBCI) state\cite{xie_fractional_2021, spanton_observation_2018, he_symmetrybroken_2023, saito_hofstadter_2021, wang_evidence_2015, waters_chern_2025}. Interestingly, the $1/3$ fraction is the only filling factor that simultaneously hosts a fractional quantum anomalous Hall state, a correlated insulator, and a SBCI at nonzero magnetic field.

\section{Thermodynamic Gap}

\begin{figure}[t]
    \centering
    \includegraphics[width= 1 \linewidth]{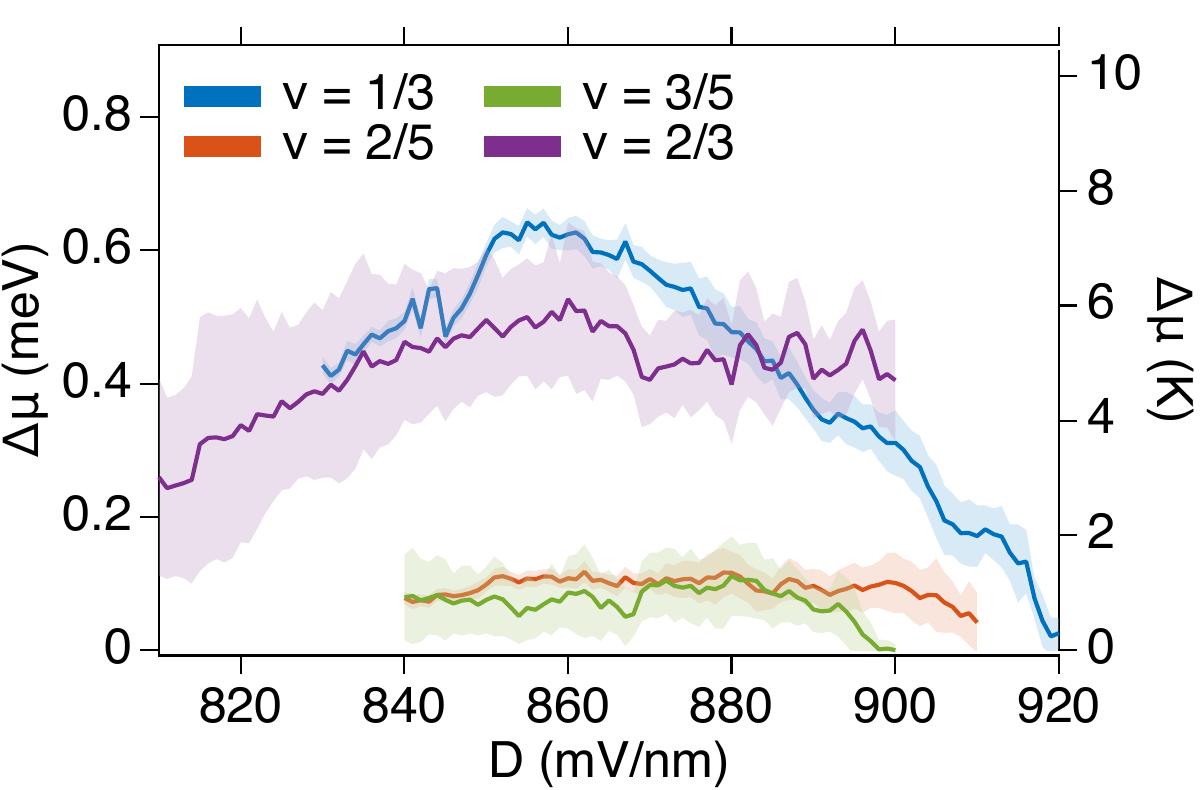}
    \caption{\justifying{\textbf{Thermodynamic Gap of FQAH States} The measured thermodynamic gap of the fractional states  as a function of displacement field at \SI{39}{\milli \kelvin}.}}
    \label{GapExtraction}
\end{figure}

From the penetration capacitance data we extract the thermodynamic gap $\Delta \mu$, the step in the chemical potential upon sweeping the electron density across a gapped state at fixed temperature (see supplementary information \ref{sec:SI_Gap_Extraction}). The size of the jump corresponds to the energy cost to adiabatically add a full electron to the system. This differs from the gap measured by thermally activated transport, $\Delta_{act}$, which determines the energy required to create a spatially separated quasiparticle-quasihole pair\cite{assouline_energy_2024}. This difference is especially important when determining the gaps of fractional quantum Hall states since the quasiparticles have fractional charge. Previous measurements of both the thermodynamic and activation gaps of fractional quantum Hall states at high magnetic field suggest that, in the clean limit, the thermodynamic gap is related to the activated gap by $\Delta\mu   = \frac{e}{e^*}\Delta_{act}$, where $e^*$ is the quasiparticle charge and $e$ is the electron charge\cite{dorozhkin_experimental_1995, eisenstein_compressibility_1994}. However, more recent studies suggest that disorder and quasiparticle interaction effects can affect the thermodynamic and activated gap in different ways, invalidating the relationship\cite{assouline_energy_2024, xie_fractional_2021, hu_highresolution_2025}.




Figure \ref{GapExtraction} shows the extracted thermodynamic gap size of the four fractional states at $\SI{39}{\milli \kelvin}$. The $1/3$ state hosts the largest gap among the FQAH states, $\Delta \mu \approx \SI{0.6}{\milli \electronvolt}$ or $\Delta \mu \approx \SI{7}{\kelvin}$. This is the largest reported gap for a graphene fractional Chern insulator. The $2/3$ gap is slightly smaller, peaking at around $\SI{0.5}{\milli \electronvolt}$ while the $2/5$ and $3/5$ gaps are much smaller. The $\nu =1$ gap peaks at $D =  \SI{770}{\milli \volt \per \nano \meter}$ with a value of $\sim \SI{2.7}{\milli \electronvolt}$ (see Fig. \ref{ExtendedData_v1_Gap}). The gap for $\nu = 1$ is similar to that reported for R$4$G/hBN\cite{choi_superconductivity_2025} and for R$5$G/hBN in the moiré distant regime with the opposite stacking orientation ($\xi = 0$)\cite{aronson_displacement_2025}. 
\begin{figure*}[t]
    \centering
    \includegraphics[width= \linewidth]{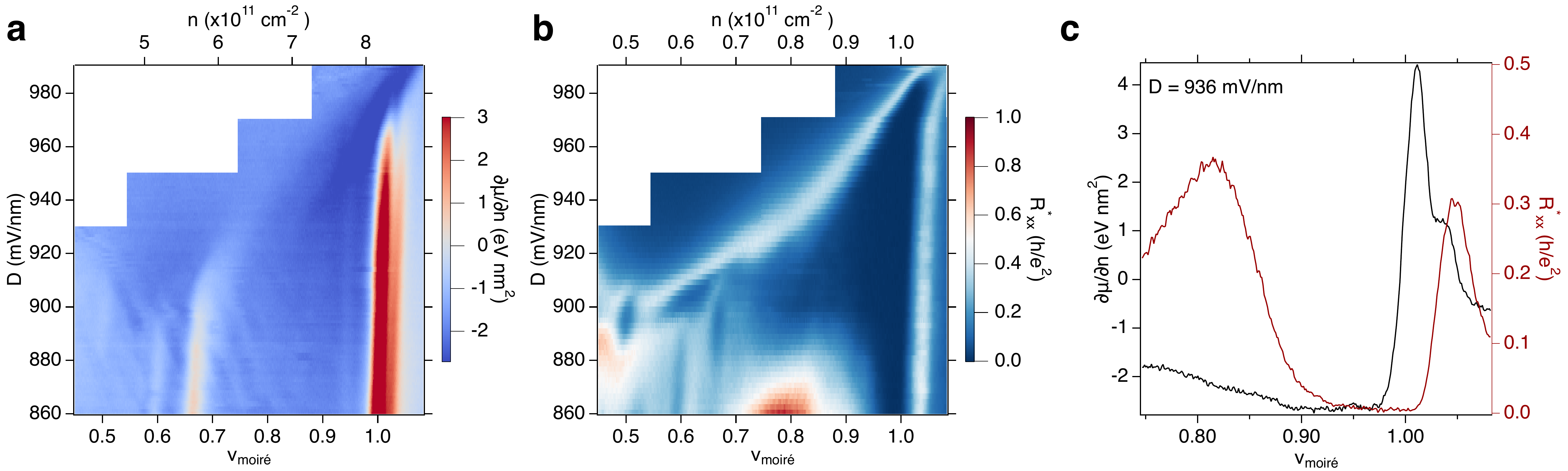}
    \caption{\justifying{\textbf{Compressibility of the Extended Quantum Anomalous Hall Regime }\textbf{a)} Compressibility and \textbf{b)} longitudinal non-local resistance ($R^*_{xx}$) measured simultaneously at $T = \SI{39}{\milli \kelvin}$ and $B = \SI{0}{\tesla}$. \textbf{c)} Line cut taken at $D = \SI{936}{\milli \volt \per \nano \meter}$} showing both the compressibility and transport. The sharp increase in $\frac{\partial \mu}{\partial n}$ occurs exactly at $\nu =1$ where transport shows a near zero $R_{xx}^*$. Note, the dramatic difference in the behaviors of the inverse compressibility and $R_{xx}^*$ in the vicinity of $\nu =1$: the inverse compressibility radically changes moving from $\nu =1$ to lower densities while $R_{xx}^*$ shows little change. At lower filling, $\nu \approx 0.9$, as the resistance begins to deviate from zero, the inverse compressibility starts to increase as well.}
    \label{EQAH}
\end{figure*}

Interestingly, both the thermodynamic gap in Fig. \ref{GapExtraction} and the phase diagram in Fig. \ref{FCI_LandauFan}\textbf{c} exhibit a striking degree of particle-hole symmetry about half-filling of the moiré band. This contrasts with the trivial CDW states, which are only present below half-filling. With the addition of the $1/3$ state, we observe the pattern of fractions as suggested by the Jain sequence, which explains the conventional FQH effect in the LLL. Figure \ref{GapExtraction} also shows a decreasing gap size as the fractions approach half-filling, which again resembles the reported behavior of the gaps in the conventional FQH effect\cite{hu_highresolution_2025, assouline_energy_2024}. 

We note the FQAH gaps that we measure are significantly smaller than the conventional FQH gaps in graphene heterostructures \cite{polshyn_quantitative_2018, assouline_energy_2024}, but more comparable to the measured gaps in gallium arsenide quantum wells\cite{du_experimental_1993, eisenstein_compressibility_1994}. One possible explanation for the smaller gap size compared to graphene systems is a proposed increased dielectric constant with increasing layer number in rhombohedral graphene\cite{xiao_density_2011}. Several other factors may contribute to the smaller gap size measured in \RGhBN including a non-zero bandwidth, disorder, and band mixing between the lowest moiré band and higher bands. Further, the gap size of the fractional states in \RGhBN is smaller than those observed in t\MoTe\cite{park_observation_2025, redekop_direct_2024}. However, this is to be expected as the densities at which the fractions occur in t\MoTe are much larger than those in \RGhBN due to a much smaller moiré wavelength.


\section{Compressibility of The Extended Quantum Anomalous Hall State}
Figure \ref{EQAH} shows both the compressibility and longitudinal resistance, $R^*_{xx}$ (see Supplementary Information \ref{sec:SI_Transport} for how $R^*_{xx}$ differs slightly from the usual $R_{xx}$), between filling $\nu \approx 1/2$ and $\nu \approx 1$. Previous very low temperature results show a vanishing longitudinal resistance along with a quantized $h/e^2$ Hall resistance over a broad region of $n-D$ phase space between $\nu = 1$ and $\nu = 1/2$\cite{lu_extended_2025}. At full-filling, $\nu =1$, we measure a large peak in the inverse compressibility, signaling a thermodynamic gap, and $R^*_{xx}$ shows a near-zero resistance, as expected for a IQAH state. However, as the system is doped to lower densities we observe a crossover from an incompressible state at $\nu = 1$ to a state with negative compressibility (i.e. very highly compressible), compared to the surrounding metallic region. Surprisingly, $R^*_{xx}$ shows no sign of a transition and remains vanishingly small throughout both the observed incompressible and compressible regions suggesting that the ballistic edge state remains unaltered across the transition. Taken together these observations indicate the EQAH state, known to display a quantized Hall resistance\cite{lu_extended_2025}, does not have a bulk thermodynamic gap. The negative compressibility and quantized Hall resistance of the EQAH state suggest a picture in which strong correlations localize charge carriers in the bulk, preventing edge state back-scattering. This picture is consistent with recent theoretical models that describe the bulk of the sample as a compressible crystalline state of either holes due to doping away from full-filling or an anomalous Hall crystal made up of electrons\cite{patri_extended_2024, dong_anomalous_2024}. 

At filling factor $\nu = 1/2$, we observe a dip in the longitudinal resistance, and previous works reported a quantized Hall resistance of $h/e^2$\cite{lu_extended_2025} at this filling. The transport properties at $\nu =1/2$ are identical to the EQAH state, previously interpreted to suggest an EQAH origin. However, the inverse compressibility shows a small peak indicating a thermodynamic gap. The presence of a gap at $\nu = 1/2$ is in sharp contrast to the negative compressibility in the EQAH region. Thus, penetration field measurements allow us to label this state as a zero-field SBCI, distinguishing it from the gapless EQAH state.

\section{Discussion}
Our measurements of R$5$G/hBN in the moiré-distant regime reveal several new features, most notably the $1/3$ FQAH state. The observation of this state motivates further theoretical work, as recent studies have predicted that the $1/3$ FQAH state is unstable to a CDW state with even minimal moiré mini-band mixing\cite{yu_moire_2025, li_multiband_2025, guo_fractional_2024}.  Further, with the observation of the $1/3$ state, both the $n-D$ phase diagram and the thermodynamic gap sizes reveal that the system displays particle-hole symmetry about half-filling, similar to Jain sequence fractions. Despite the proposed unconventional origin of the topology in the system\cite{tan_ideal_2025, li_multiband_2025, yu_moire_2025, huang_fractional_2025, zhou_fractional_2024, dong_theory_2024, huang_selfconsistent_2024, guo_fractional_2024, xie_integer_2024, song_intertwined_2024}, the fractional states appear very similar to those observed in the LLL at high magnetic field.


Our compressibility and transport measurements show a crossover between a gapped IQAH state at $\nu =1$ to a gapless EQAH state at lower densities that displays strong negative compressibility but retains a ballistic edge mode over an extended range in density. A bulk ordered state with a ballistic edge mode is thought to explain the integer quantum Hall Wigner solid\cite{jang_sharp_2017,chen_microwave_2003, lewis_wigner_2004, zhou_solids_2020} and also the re-entrant integer quantum Hall effect\cite{lilly_evidence_1999, fogler_ground_1996, koulakov_charge_1996, eisenstein_insulating_2002, lewis_microwave_2005, deng_collective_2012, chen_competing_2019, zhou_solids_2020}. A similar behavior may give rise to RIQAH states in t\MoTe\cite{xu_signatures_2025}. However, direct observation of a transition from an incompressible state to a compressible state while retaining a topological edge state has not previously been observed in any system. More direct measurements of electron crystallinity are needed and could come from narrow-band noise at microwave frequencies \cite{gruner_observation_1981, fleming_slidingmode_1979}, spatially resolved probes such as scanning tunneling microscopy\cite{xiang_imaging_2025, tsui_direct_2024}, or planar tunneling spectroscopy\cite{jang_sharp_2017}.

Taken together, our measurements suggest that the fractional states strongly resemble those of the Jain sequence in the LLL\cite{jain_compositefermion_1989}, despite a fundamentally different physical origin. This calls for new theoretical models that can account for the similarity of the FQAH states to the fractional states in the LLL.

\putbib
\end{bibunit}
\newpage
\clearpage
\begin{bibunit}
\section{Methods}
\subsection{Device Fabrication}
The graphene and hBN flakes were prepared by mechanical exfoliation onto SiO$_2$/Si substrates. The rhombohedral domains of pentalayer graphene were identified and confirmed using IR camera, near-field infrared microscopy, and Raman spectroscopy and isolated by cutting with a femtosecond laser. The van der Waals heterostructure was made following a dry transfer procedure. We picked up the top hBN, graphite, middle hBN and graphene using poly(bisphenol A carbonate) film on polydimethylsiloxane, and landed it on a prepared bottom stack consisting of an hBN and graphite bottom gate. The device was then etched into a multiterminal structure using e-beam lithography and reactive-ion etching. We deposited Cr–Au for electrical connections to the source, drain and gate electrodes.

\subsection{AC Capacitance Measurements}\label{sec:Capacitance_Methods}
All capacitance measurements on device D1 at $\SI{320}{\milli \kelvin}$ or above were carried out in a Oxford Heliox He$^3$ refrigerator. To measure the penetration capacitance of the device we utilize a capacitance bridge circuit as shown in Fig. \ref{SI_He3_Circuit}\cite{aronson_displacement_2025, delabarrera_cascade_2022, samuelh.aronson_electronic_2025, spencerlouistomarken_thermodynamic_2019}. A small excitation is applied to the top gate of the device. A second excitation is then applied to the standard reference capacitor with capacitance of approximately $\SI{25}{\femto \farad}$ to null the signal at the input of a cryogenic amplifier. As we vary the gate voltages or the applied magnetic field, the bridge output moves away from null, and we compute the device impedance based on the measured off-balance signal\cite{raymondc.ashoori_density_1991}. To measure the capacitance of the small graphene device, we place a cryogenic trans-impedance amplifier within a few millimeters of the device. The signal is subsequently amplified again by a room temperature amplifier before being measured by a lock-in amplifier (See supplementary information Section \ref{sec:SI_Circuit_Diagram}). We use an excitation of $V_{rms} = \SI{2.37}{\milli \volt}$ for all measurements at $T = \SI{320}{\milli \kelvin}$ and an excitation of  $V_{\textrm{rms}} = \SI{7}{\milli \volt}$ for all measurements at $T = \SI{39}{\milli \kelvin}$ unless otherwise noted (See \ref{SI_excDependence} for details on excitation dependence).

In the low-frequency limit wherein the device can fully charge on an AC time scale, the penetration capacitance per unit area can be written as $c_P^{-1} = c_{tg}^{-1} + c_{bg}^{-1} + \frac{\partial n}{\partial \mu} \frac{e^2}{c_{tg} c_{bg}}$, making penetration capacitance, $c_P$, a sensitive probe of the compressibility, $\frac{\partial n}{\partial \mu}$, in the graphene system. Here, $c_{tg}$ ($c_{bg}$) is the geometric capacitance per unit area between the graphene and the top gate (back gate). When the device has a sufficiently large in-plane resistance, the graphene layer cannot charge and discharge on an $AC$ timescale. Thus, in this limit the penetration capacitance is now a frequency dependent measurement of both the density of states and the in plane resistance. This limit is also indicated by a large dissipative, or out of phase signal in this region (see Fig. \ref{Extended_Data_Closs_PhaseDiagram}). Unlike transport measurements, which struggle to measure very large resistances, capacitance is sensitive to resistances that affect the $R C$ charging time of the device. For small van der Waals devices this is typically on the order of $\SI{ 1}{\giga \ohm}$. This makes high-frequency capacitance measurements sensitive to changes in very large resistance values. This enables us to resolve fine features within the WS region (see supplementary information).

When measuring at $\SI{320}{\milli \kelvin}$, we control the density using $V_{tg}$ and $V_c$. $V_{tg}$ is the voltage applied to the top-gate and $V_c$ is the voltage applied to the graphene flake. The carrier density is then set by, $n = (c_{tg} V^*_{tg} +c_{bg} V^*_{bg})/e +n_{off}$ where $V^*_{bg} = -V_c$ and $V^*_{tg} = V_{tg}-V_c$. $n_{off}$ is chosen such that the correlated insulator at $D = \SI{0}{\milli \volt \per \nano \meter}$ is centered at $n = \SI{0}{\per \cm \squared}$. The displacement field is calculated as $D = e(c_{tg} V^*_{tg} - c_{bg} V^*_{bg})/2 \varepsilon_0 + D_{off}$. $D_{off}$ is chosen such that the correlated insulator at $D = \SI{0}{\milli \volt \per \nano \meter}$ is centered at zero displacement field. Both $c_{tg}$ and $c_{bg}$ were determined by fitting the spacing between Landau levels at $D = \SI{0}{\milli \volt \per \nano \meter}$. 

Capacitance and transport data for device D1 were measured at $\SI{39}{\milli \kelvin}$ in a Bluefors LD250 dilution refrigerator. In this setup, we place a $\SI{2}{\pico \farad}$ decoupling capacitor between the device and the cryogenic amplifier to minimize device heating from the amplifier. In this setup we use top and bottom gate voltages to control the carrier density, $n = (c_{tg} V_{tg} +  c_{bg} V_{bg})/e + n_{off}$ and displacement, $D =(c_{tg} V_{tg} - c_{bg} V_{bg})/2 \epsilon_0 + D_{off}$ (See supplementary information \ref{sec:SI_Dilution_CircuitDiagram} for further details.).

\subsection{Transport Measurements}
We measure the transport of device D1 simultaneously with the capacitance in the dilution refrigerator using standard lock-in transport techniques. 
We current bias the device by applying $\SI{30}{\milli \volt}$, $\SI{13.15}{\hertz}$, AC excitation in series with a $\SI{300}{\mega \ohm}$ bias resistor to create a $\SI{100}{\pico \ampere}$ current bias. We keep the device drain contact grounded at low temperature (See supplemental information \ref{sec:SI_Transport} for transport measurement schematic). A Basel Precision Instruments preamplifier was used to measure the differential voltage. 
    
Device D2 was measured in a Bluefors LD250 dilution refrigerator with a minimum electronic temperature of around $\SI{40}{\milli \kelvin}$. Stanford Research Systems SR830 and SR860 lock-in amplifiers and SP1004 voltage preamplifiers from Basel Precision Instruments were used to measure the longitudinal and Hall resistance $R_{xx}$ and $R_{xy}$ with an AC frequency at $\SI{17.77}{\hertz}$. The AC currents are generated by the SR860 lock-in through a $\SI{10}{\mega \ohm}$ resistor. The AC current excitation was limited to be $\SI{1}{\nano \ampere}$ or below. Basel Precision Instruments preamps were used to measure differential currents and voltages. Keithley 2400 source-meters were used to apply top and bottom gate voltages. The top-gate voltage $V_{tg}$ and the bottom-gate voltage $V_{bg}$ are swept to adjust the carrier density $n=(c_{tg}V_{tg}+c_{bg}V_{bg})/e$ and displacement field $D=(c_{tg}V_{tg}-c_{bg}V_{bg})/2\varepsilon_0$, where $c_{tg}$ and $c_{bg}$ are top-gate and bottom-gate capacitances per area calculated from the Landau fan diagram.

Device D2 is a modified Hall bar. We measure a resistance $R_{mix}$  which contains both $R_{xx}$ and $R_{xy}$ components. $R_{xx}$ ($R_{xy}$) is extracted from symmetrizing (anti-symmetrizing) the mixed resistance signal with respect to positive and negative magnetic field;  $R_{xx}=(R_{mix}(+B)+R_{mix}(-B))/2$ and $R_{xy}=(R_{mix}(+B)-R_{mix}(-B))/2$ . $\pm \SI{0.1}{\tesla}$  is used for this procedure unless otherwise stated.

\section{Acknowledgments}
We acknowledge helpful discussions with Aidan Reddy, Jonah Herzog-Arbeitman, and Patrick Ledwith. Work from the Ashoori group (J.B., A.D., and R.A.) was supported by the U.S. Department of Energy, Office of Science, Basic Energy Sciences (ECMP), under Award No. DE-SC0026083. The device fabrication and DC transport measurement parts were supported by the US Department of Energy, Office of Science, Basic Energy Sciences, Materials Sciences and Engineering Division under grant no. DE-SC0025325. Device fabrication was performed at the Harvard Center for Nanoscale Systems and MIT.nano. J.B. and T.H. acknowledge support from a Mathworks Fellowship. K.W. and T.T. acknowledge support from the JSPS KAKENHI (grant nos. 20H00354, 21H05233 and 23H02052) and the World Premier International Research Center Initiative, MEXT, Japan. A.S. acknowledges support from the ISF, ISF Quantum Science and Technology (grant 2074/19) and the DFG (CRC/Transregio 183).

\putbib
\end{bibunit}

\newpage
\clearpage

\newpage
\clearpage

\counterwithin{figure}{section}
\renewcommand{\thefigure}{Extended \arabic{figure}}
\setcounter{figure}{0}
\newpage
\clearpage


    \begin{figure*}[t]
        \begin{center}
            \LARGE\textbf{Extended Data}
        \end{center}
        \includegraphics[width = 0.9 \linewidth]{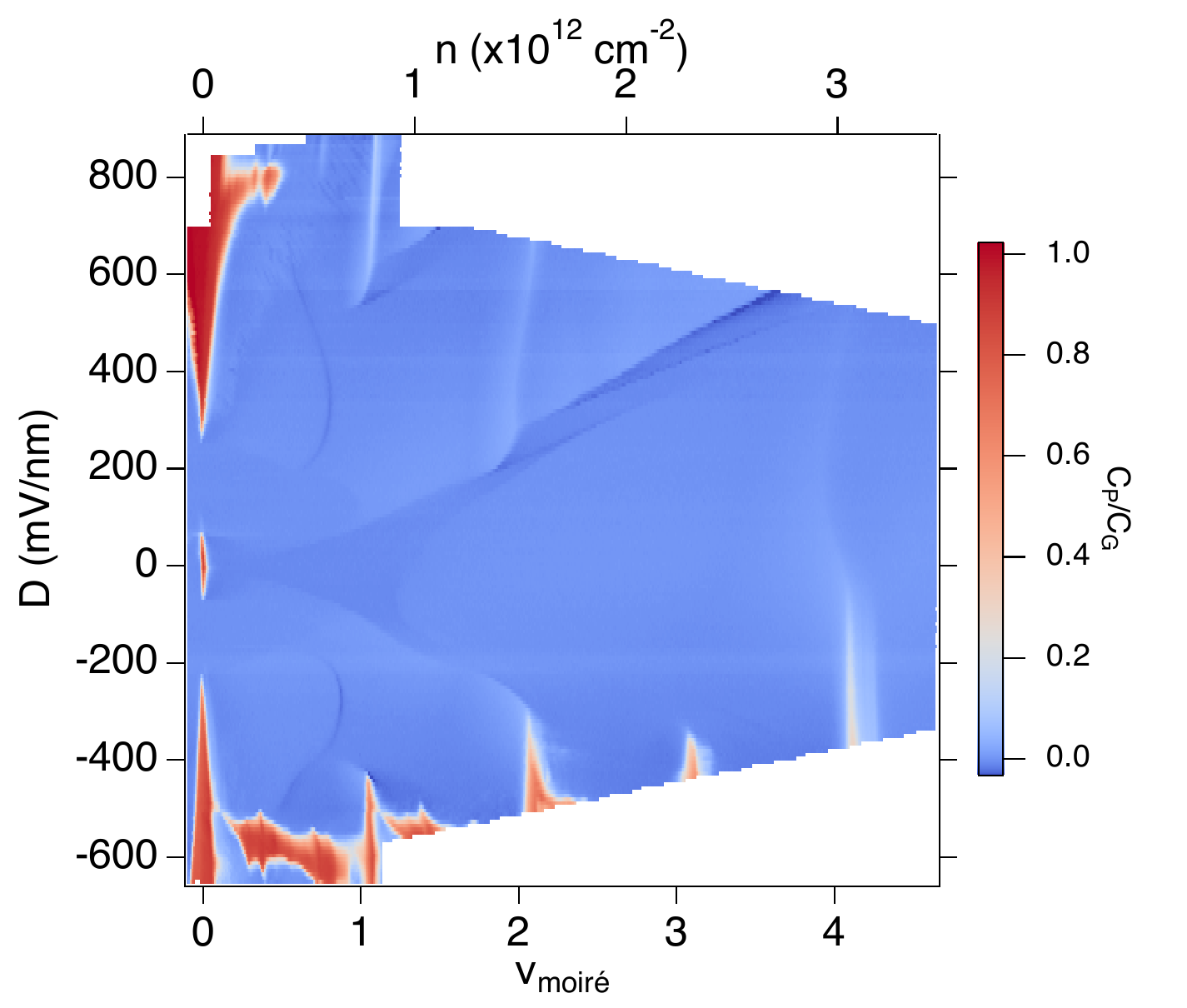}
        \caption{\justifying{Full phase diagram as measured by penetration capacitance at $f = \SI{3.78}{\mega \hertz}$ and $T = \SI{320}{\milli \kelvin}$. We identify the stacking order as $\xi =1$ due to the presence of incompressible states at $\nu =3 \textrm{ and } 4$ in the moiré-proximal regime, negative displacement field,  according to ref. \cite{uzan_hbn_2025}. Note, any observed horizontal banding arises from  cryogenic amplifier shifts.}}
        \label{Extended_Data_FullPhaseDiagram}
    \end{figure*}

    \begin{figure*}[t]
        \centering
        \includegraphics[width = 0.9 \linewidth]{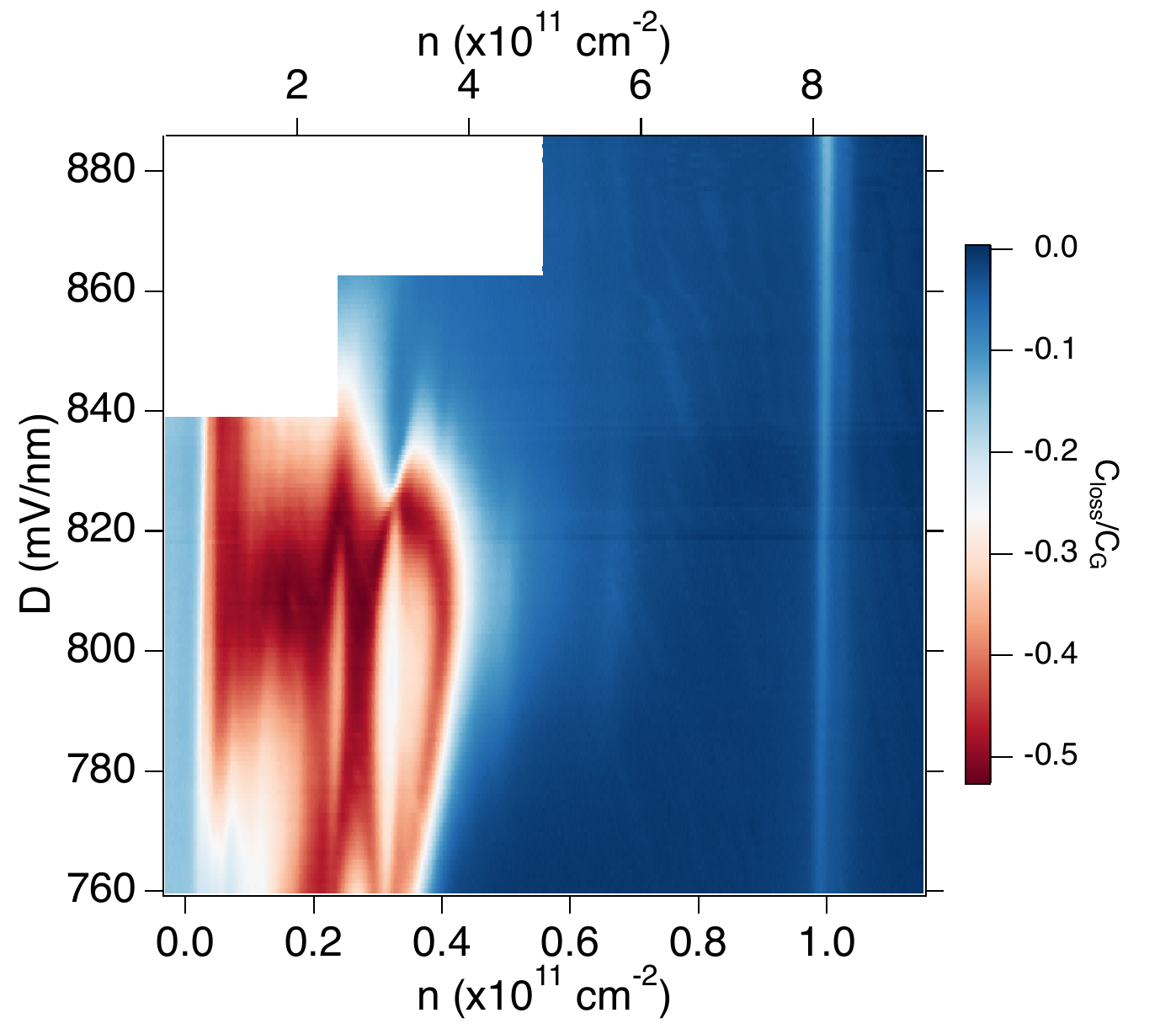}
        \caption{\justifying{C$_\textrm{loss}$ or the dissipative component of the capacitance measurement shown in Fig. \ref{PhaseDiagram}\textbf{a}. This component arises from a high in plane resistance of the sample. We see a negligible $C_{\textrm{loss}}$ signal associated with the FQAH states, but a substantial dissipative signal from the WS region.}}
        \label{Extended_Data_Closs_PhaseDiagram}
    \end{figure*}
\newpage 
    \begin{figure*}[t]
        \includegraphics[width = 0.9 \linewidth]{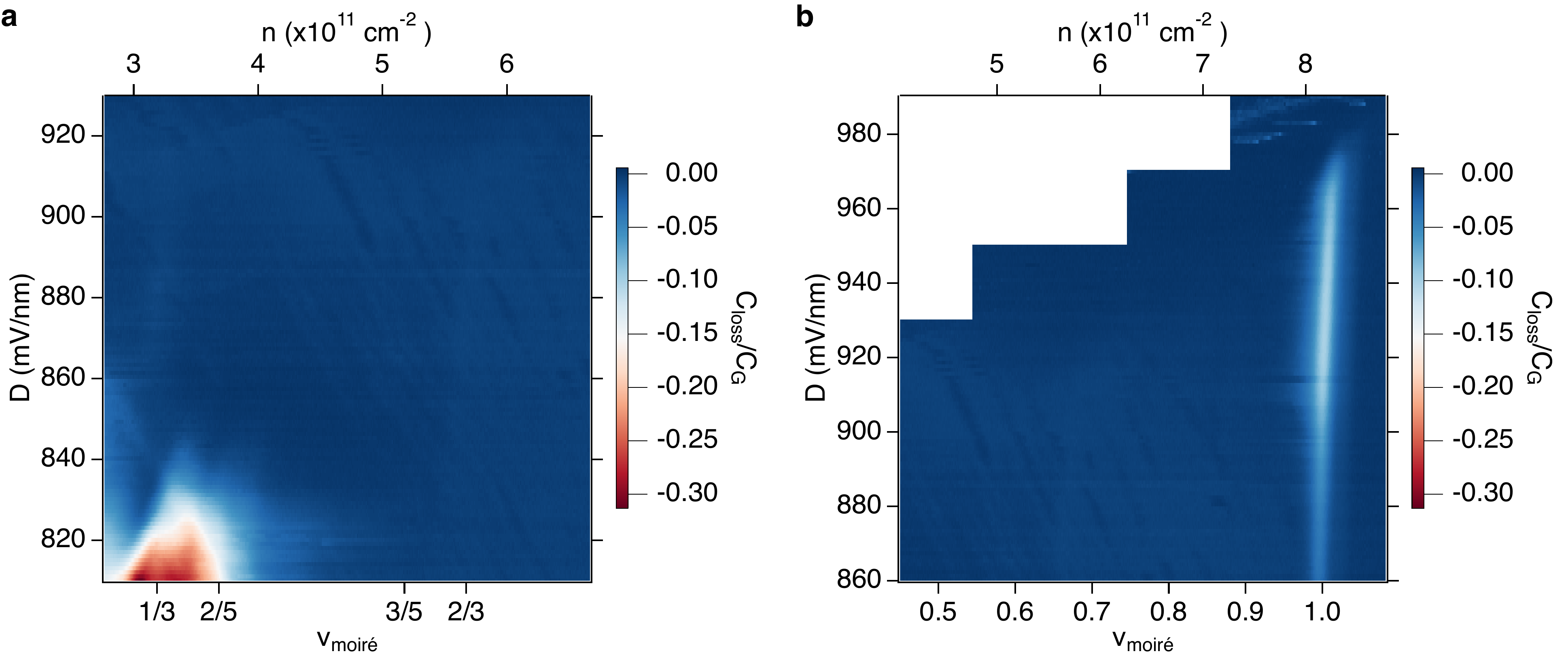}
        \caption{\justifying{C$_\textrm{loss}$ or the dissipative component of the capacitance measurement as shown in Fig. \ref{FCI_LandauFan}\textbf{a} and \ref{EQAH}\textbf{a}}}
        \label{Extended_Data_Closs_DilutionFridge}

    \end{figure*}
\newpage
\begin{figure*}[t]
    \centering
    \includegraphics[width=0.9 \linewidth]{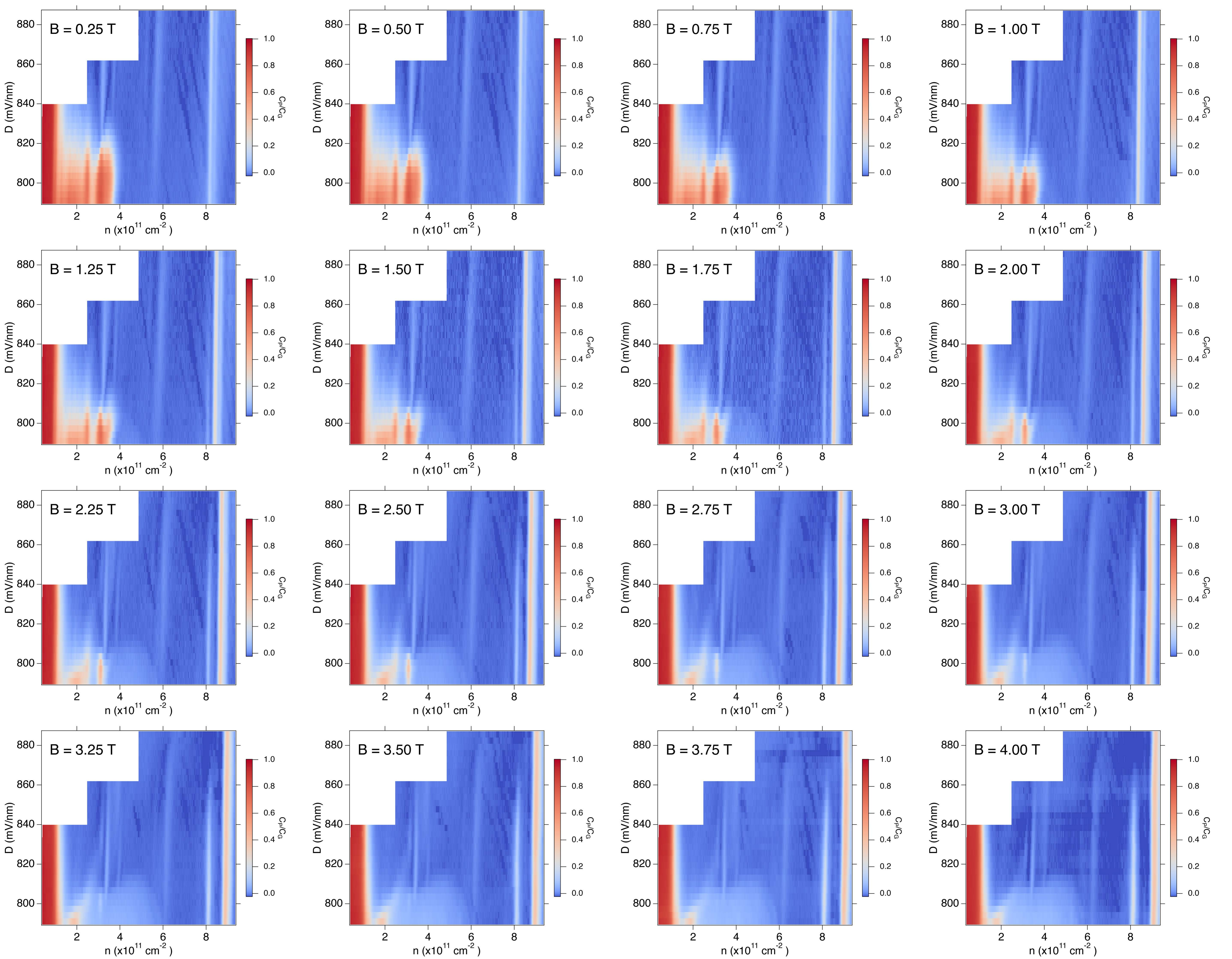}
    \caption{\justifying{Phase diagram as measured by penetration capacitance at a series of magnetic fields. Capacitance measurements took place at $\SI{320}{\milli \kelvin}$ and $f = \SI{10}{\mega \hertz}$. The FQAH states move in density with the expected magnetic field dependence, while the CDW states do not move in density. }}
    \label{nvsDvsB_maps}
\end{figure*}
\newpage

\begin{figure*}[t]
    \centering
    \includegraphics[width = 1 \linewidth]{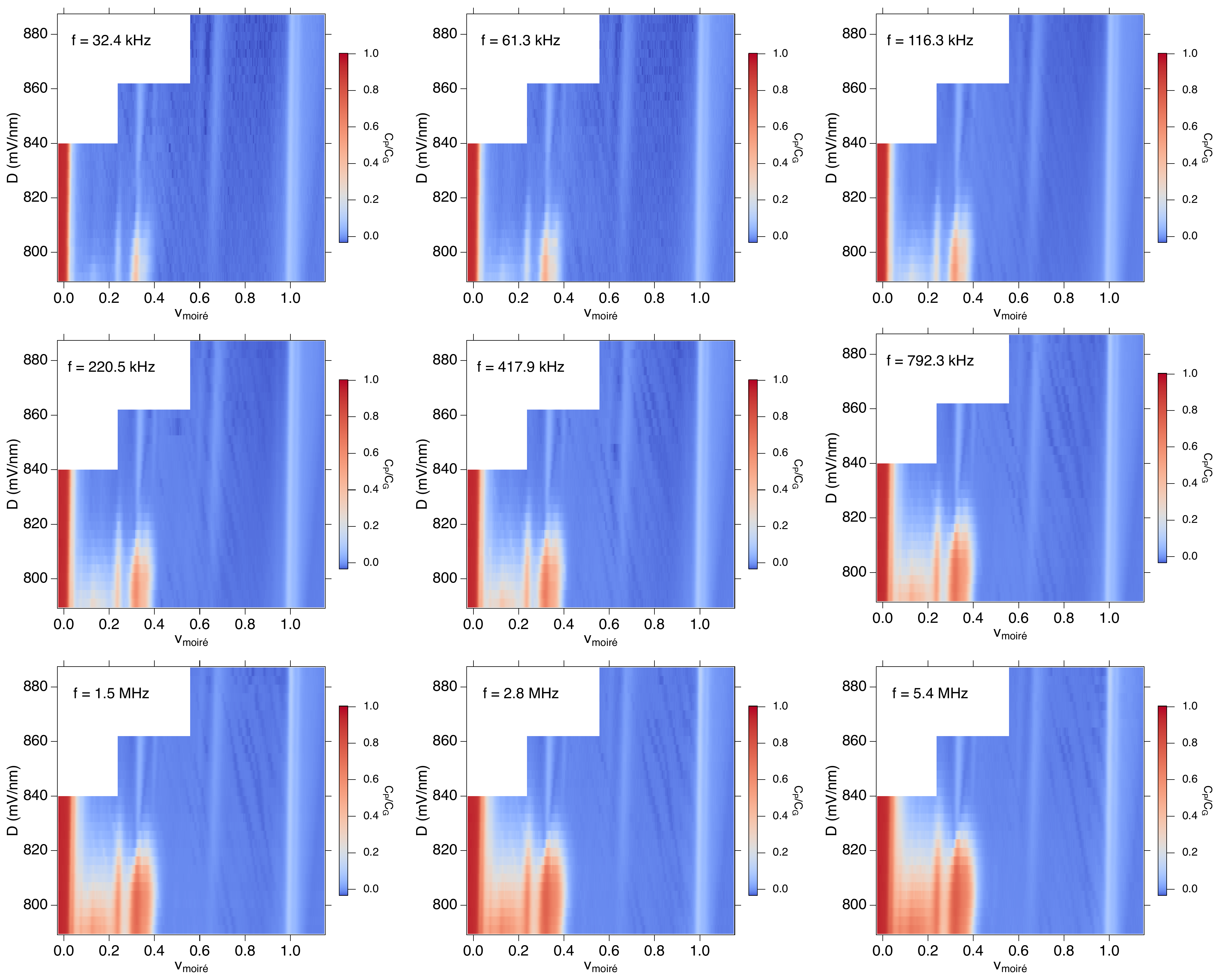}
    \caption{\justifying{Phase diagram as measured by penetration capacitance across a range of different frequencies at $T = \SI{320}{\milli \kelvin}$. As the frequency is increased $C_P$ increases in the WS region as is expected for states with very high in plane resistances.}}
    \label{ExtendedData_nvdDvdf_maps}
\end{figure*}

\begin{figure*}[t]
    \centering
    \includegraphics[width = \linewidth]{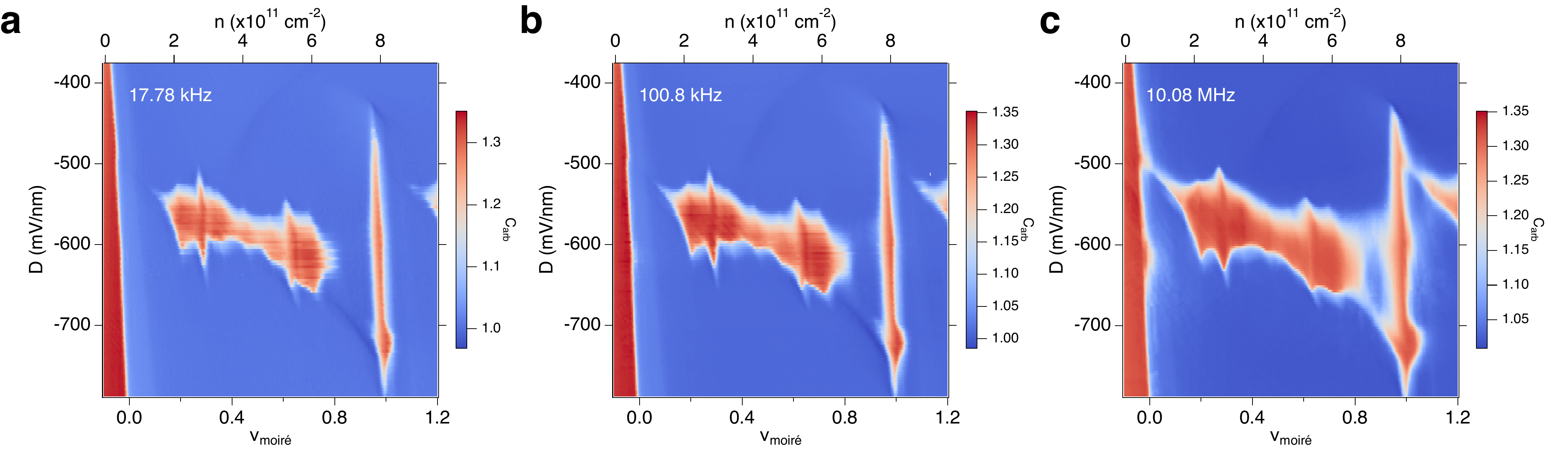}
    \caption{\justifying{Penetration capacitance as measured on the moire proximal side at zero magnetic field measured at \textbf{a)} $\SI{17.78}{\kilo\hertz}$, \textbf{b)} $\SI{100.8}{\kilo \hertz}  $  and \textbf{c)} $\SI{10.08}{\mega \hertz}$ at base temperature of $T = \SI{320}{\milli \kelvin}$. Capacitance here is reported in arbitrary units.}}
    \label{ExtendedData_MoireProximal}
\end{figure*}

\begin{figure*}[t]
    \centering
    \includegraphics[width=0.95\linewidth]{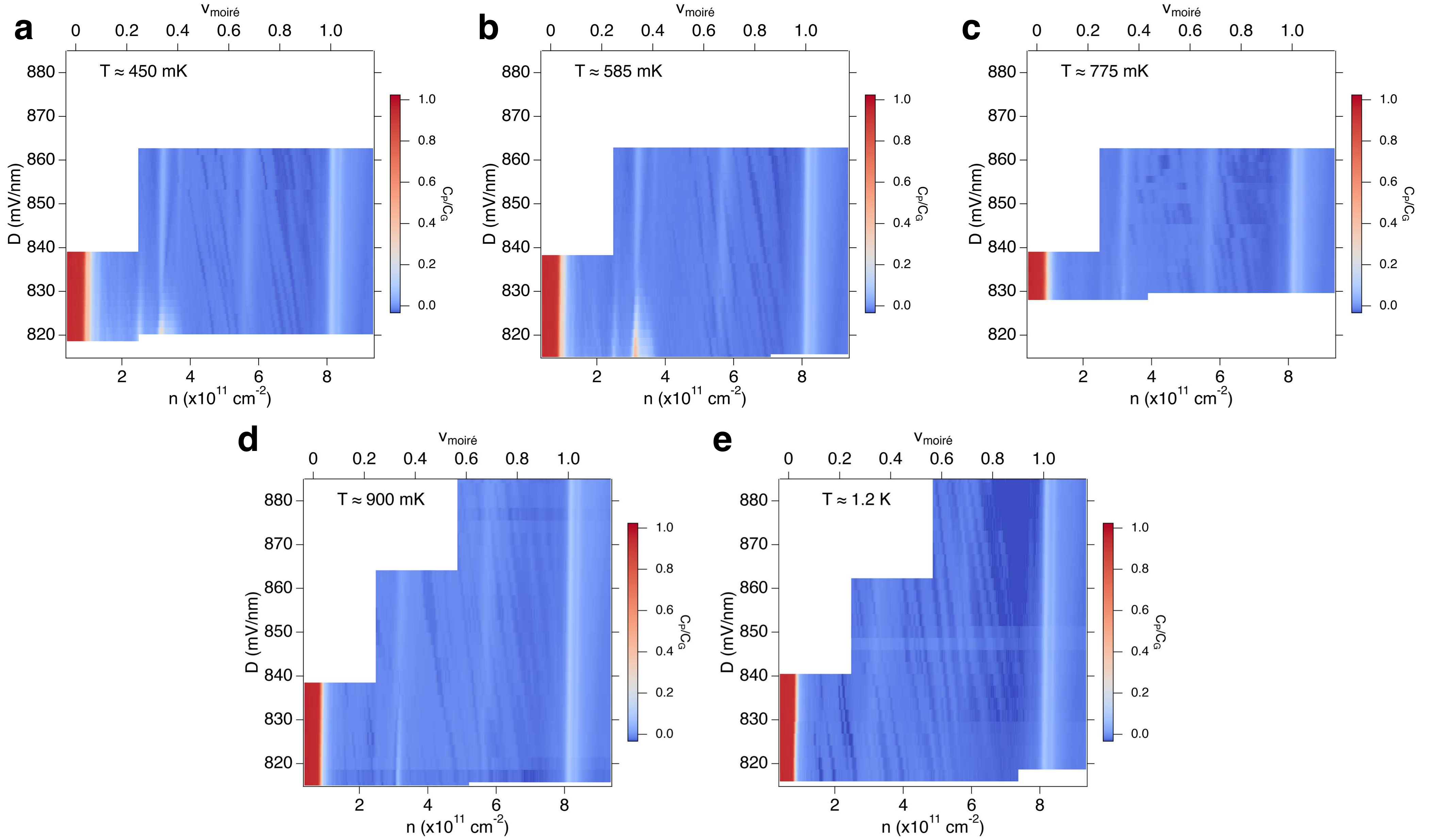}
    \caption{\justifying{n-D maps taken at different temperatures. All maps taken at a frequency of $f = \SI{1}{\mega \hertz}$. We observe that the diagonal contact features become enhanced as a function of temperature (see supplement section IIc).}}
    \label{ExtendedData_TempDependence}
\end{figure*}

\begin{figure*}[t]
    \centering
    \includegraphics[width = 0.5 \linewidth]{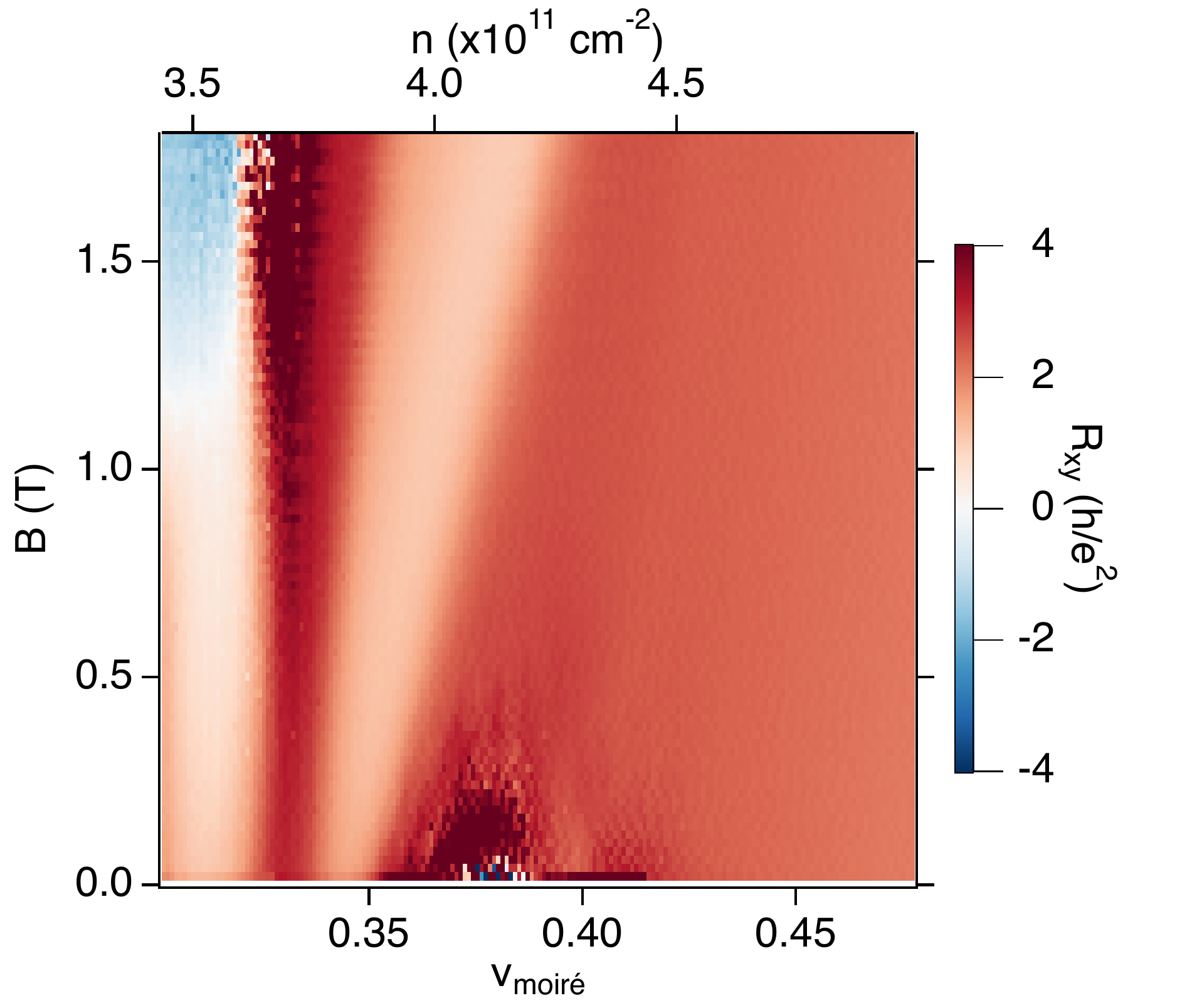}
    \caption{\justifying{Landau fan measurement display $R_{xy}$ measurement for device \textbf{D2}. Data has been anti-symmetrized.}}
    \label{ExtendedData_RxyLandauFan}
\end{figure*}

\begin{figure*}[t]
    \centering
    \includegraphics[width = 0.9 \linewidth]{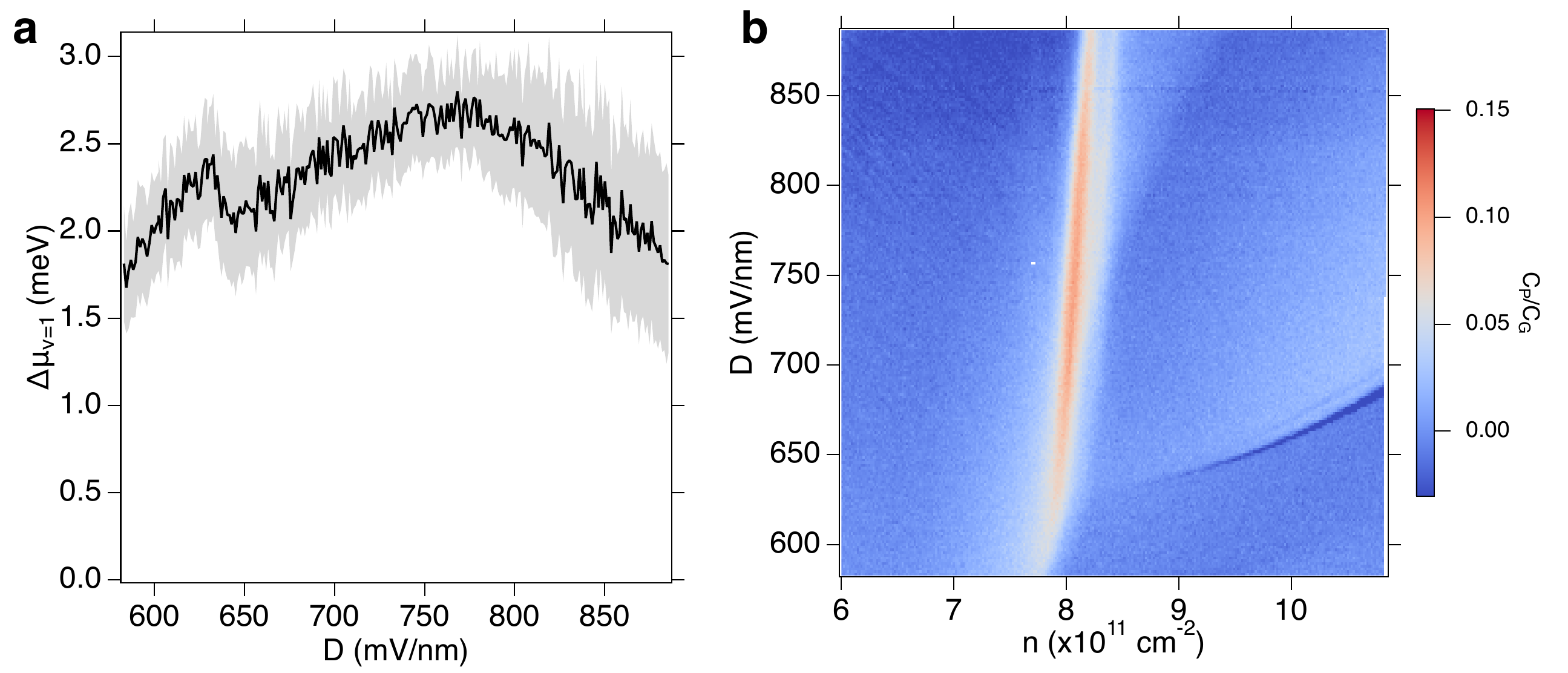}
    \caption{\justifying{Capacitance of $\nu =1$ state a function of $n-D$ at $T = \SI{320}{\milli \kelvin}$ and $f = \SI{38}{\kilo \hertz}$ with an excitation of $V_{rms} = \SI{0.95}{\milli \volt}$}.}
    \label{ExtendedData_v1_Gap}
\end{figure*}

\newpage
\clearpage
\begin{bibunit}
    
\renewcommand{\thefigure}{SI \arabic{figure}}

\newcommand{\mvnm}{\milli \volt \per \nano \meter}

\title{Supplementary Information}
\maketitle
\onecolumngrid  
\section{Capacitance Circuit Diagram}\label{sec:SI_Circuit_Diagram}

\begin{figure*}[h]
        \includegraphics[width = 0.5\linewidth]{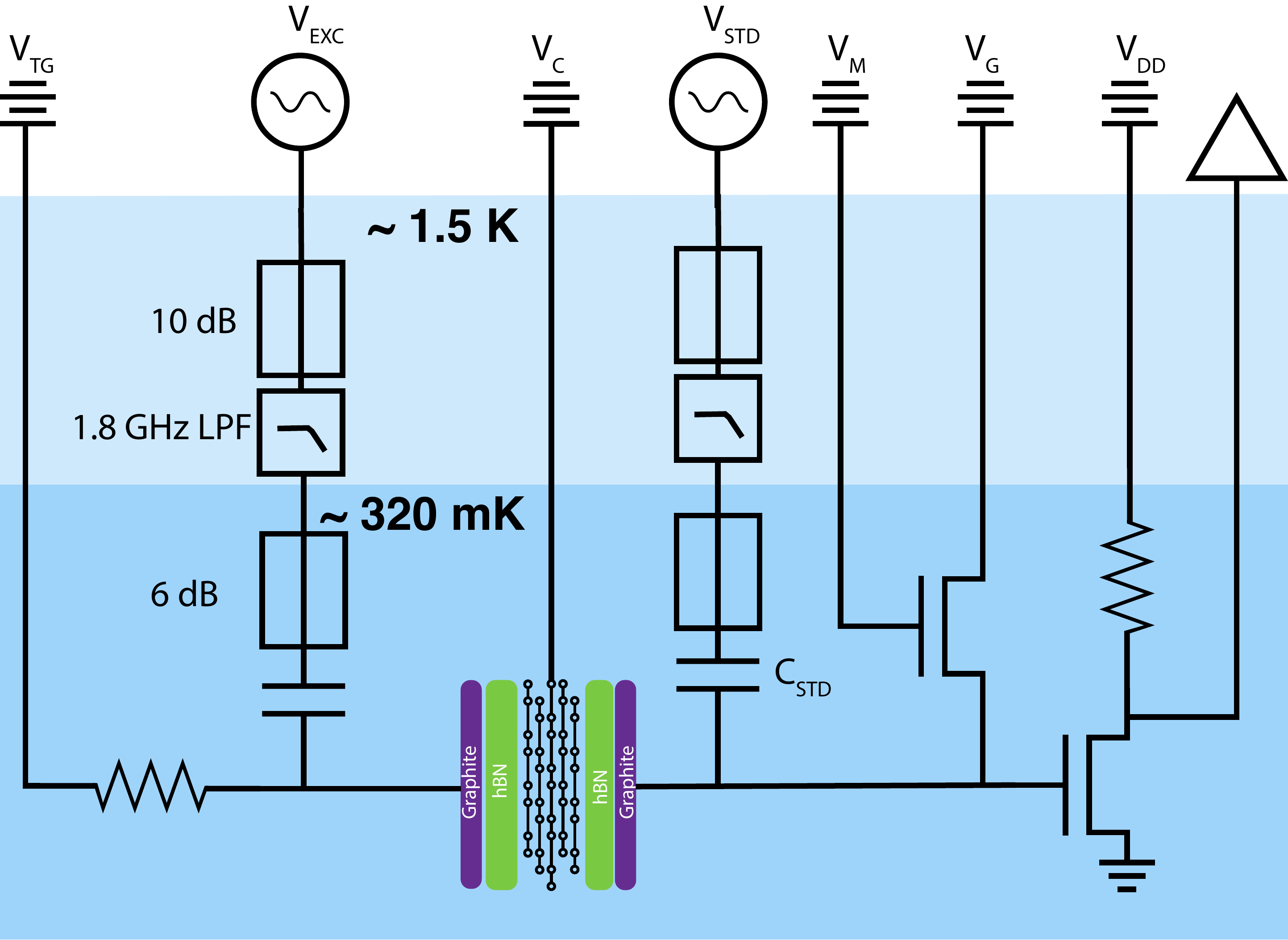}
        \caption{\justifying{Circuit Diagram used to measure the penetration capacitance in a He$^3$ refrigerator. The AC excitations and signals are run through semi-rigid coaxial cables. The transistors shown are FHX35X HEMTs. Additional filtering applied to the DC lines is not shown.}}
        \label{SI_He3_Circuit}
    \end{figure*}

As described in the methods section of the main text we use a capacitance bridge circuit as shown in Fig. \ref{SI_He3_Circuit} and \ref{SI_DF_Circuit}. 

    \subsection{\SI{320}{\milli \kelvin} Capacitance Measurements}

\subsection{Dilution Refrigerator Capacitance Measurements}\label{sec:SI_Dilution_CircuitDiagram}
\begin{figure*}[h]
        \includegraphics[width = 0.5 \linewidth]{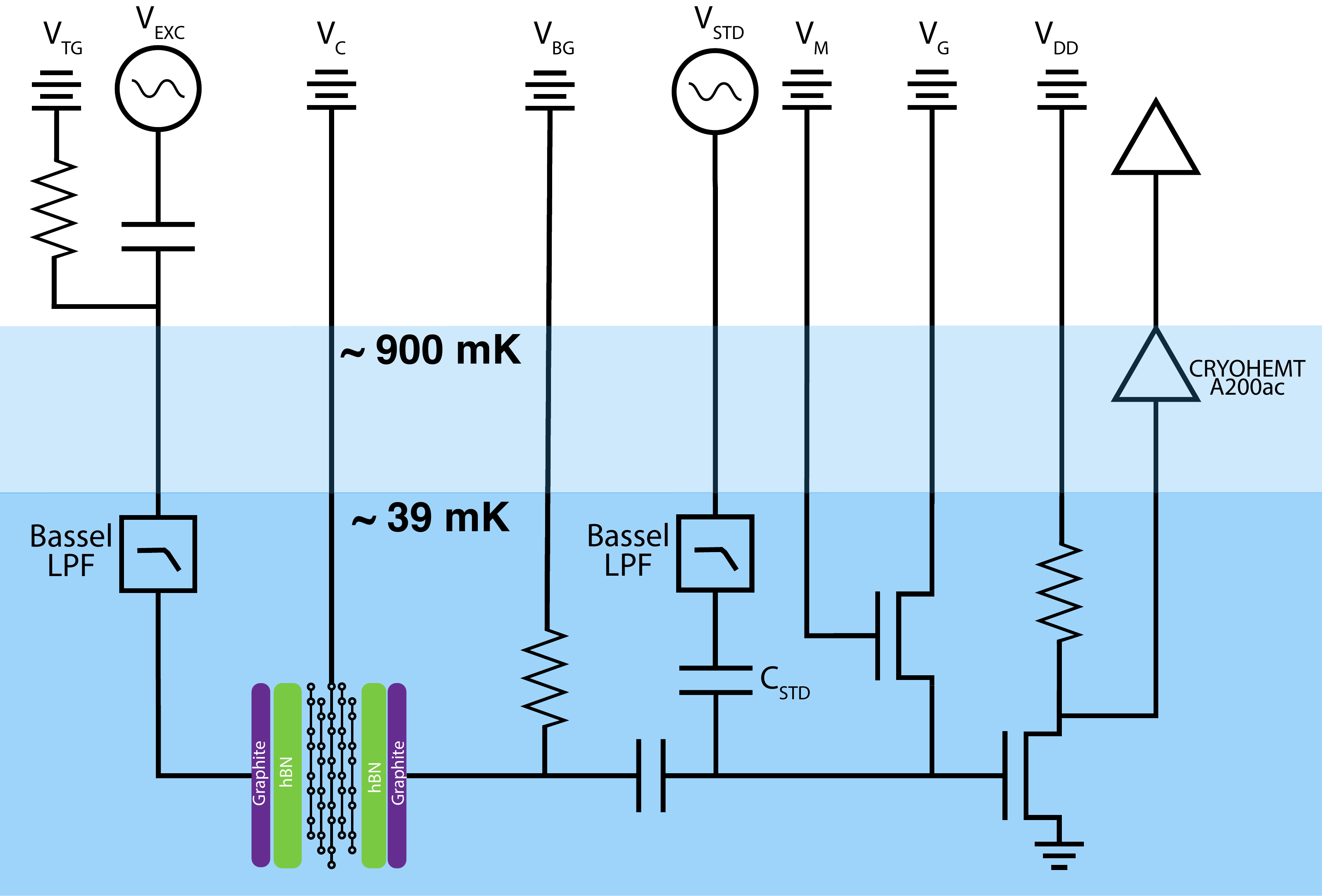}
        \caption{\justifying{Circuit used to measure the penetration capacitance in a BlueFors LD250 Dilution Refrigerator. The transistors shown or FHX35X HEMTs. Additional filtering to the DC lines Additional filtering applied to the DC lines as well as room temperature electronics is not shown. Only filters and attenuators placed inside the fridge on the AC lines is shown.}}
        \label{SI_DF_Circuit}
    \end{figure*}

To measure the penetration capacitance of the sample at dilution refrigerator temperatures, we make some alterations to the capacitance circuit used in the He$^3$ refrigerator. As shown in Fig. \ref{SI_DF_Circuit}, we place a capacitor between the gate of the sample and the amplifier in order to reduce heating of the sample caused by the amplifier. We use a CRYOHEMT A200ac amplifier as a second stage low temperature amplifier placed at the still plate of the dilution refrigerator. This allows us run the first stage cryogenic amplifier at a very low power to reduce heating effects. We make simultaneous measurements of transport and capacitance to ensure that the amplifier power does not drastically alter the transport characteristics of the sample.

\section{Compressibility Normalization and Gap Extraction}\label{sec:SI_Compressibility_Normalization}
As shown in the main text, penetration capacitance can be related to the thermodynamic compressibility through the relation  $c_P^{-1} = c_{\textrm{tg}}^{-1} + c_{\textrm{bg}}^{-1} + \frac{\partial n}{\partial \mu} \frac{e^2}{c_{\textrm{tg}} c_{\textrm{bg}}}$. In this formula, the $c$'s are capacitances per unit area (denoted by letter-case $c$, whereas absolute capacitances are denoted capital letters). We use a bridge setup to measure the penetration capacitance of the sample. We measure the ratio, $\chi$, of the sample penetration capacitance to a reference capacitor $C_{std}$, which is approximately $\SI{25}{\femto \farad}$. To convert from $C_P/C_{std} \equiv \chi$ to compressibility, we use a similar procedure to that used for gap extraction in the supplement of ref.  \cite{aronson_displacement_2025}. We convert this to compressibility using the expression:  

$$\frac{\partial \mu}{\partial n} = e^2 \frac{\chi - \chi_{\textrm{band}}}{\chi_{\textrm{gap}}- \chi} \frac{1}{c_{\textrm{tg}} + c_{\textrm{bg}}}$$

Here, $\chi_{\textrm{band}}$ is $\chi$ measured in the band, and $\chi_{\textrm{gap}}$ is $\chi$ measured in the gap. As discussed in, \cite{aronson_displacement_2025} choosing $\chi_{\textrm{band}}$ can be difficult. Ideally we chose a point in the phase space where $\frac{\partial \mu}{\partial n}\rightarrow 0$.  We note that the choice of background only negligibly impacts the gap measurement. The effect of the background choice appears mainly as a shift in the zero of the inverse compressibility.

\subsection{Gap Extraction}\label{sec:SI_Gap_Extraction}
Once the capacitance ratio measured is converted to compressibility, we can subsequently use the compressibility to compute the thermodynamic gap in the same way as ref. \cite{aronson_displacement_2025} using the relation $ c_{\textrm{q}}^{-1} = e^2 \frac{\partial \mu}{\partial n}$. Then we compute the gap by taking the integral, 

$$\Delta \mu = e \int dV_{\textrm{tg}} (1+\frac{c_{\textrm{q}}}{c_{\textrm{tg}}})^{-1} +  e \int dV_{\textrm{bg}} (1+\frac{c_{\textrm{q}}}{c_{\textrm{bg}}})^{-1} =  e \int dV_{\textrm{tg}} m_{\textrm{t}}(c_{\textrm{q}}, c_{\textrm{tg}}) +  e \int dV_{\textrm{bg}} m_{\textrm{b}}(c_{\textrm{q}}, c_{\textrm{bg}}) $$

Here we define, $m_{\textrm{t}} = c_{\textrm{q}}^{-1}(\frac{1}{c_{\textrm{tg}}} + \frac{1}{c_{\textrm{q}}})^{-1}\textrm{, } m_{\textrm{b}} = c_{\textrm{q}}^{-1}(\frac{1}{c_{\textrm{bg}}} + \frac{1}{c_{\textrm{q}}})^{-1}$

Since our data set consists of discrete points, we compute the gap using:

$$\Delta \mu = e \sum_i \left(1+\frac{c_{\textrm{q}}^{(i)}}{c_{\textrm{tg}}}\right)^{-1} \Delta V_{\textrm{tg}}^{(i)} + e \sum_i \left(1+\frac{c_{\textrm{q}}^{(i)}}{c_{\textrm{bg}}}\right)^{-1} \Delta V_{\textrm{bg}}^{(i)}$$

\begin{figure}
    \centering
    \includegraphics[width=0.9\linewidth]{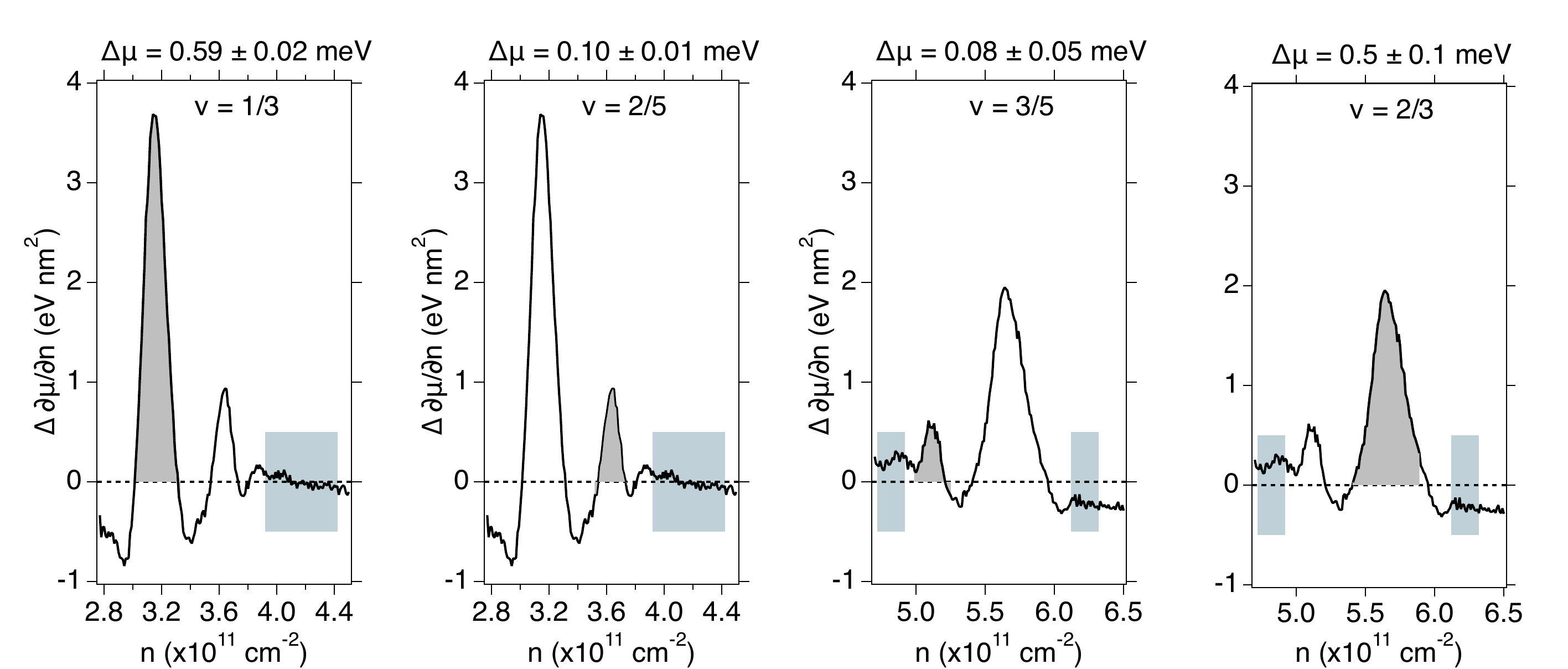}
    \caption{\justifying{Compressibility trace at $D= \SI{850}{\milli \volt \per \nano \meter}$ and $T = \SI{39}{\milli \kelvin}$. Gray area under the curve represents the area integrated to get the thermodynamic gap. The blue shaded region is the neighboring region which is taken to be $\chi_{\textrm{band}}$.}}
    \label{SI_GapExample}
\end{figure}

As described in \cite{aronson_displacement_2025, samuelh.aronson_electronic_2025}, determining $\chi_{\textrm{band}}$ is difficult as the fractional states reside on a background of non-zero inverse compressibility. Thus, we take $\chi_{\textrm{band}}$ to be a flat region near the fractional state, as shown in Fig. \ref{SI_GapExample}. We then use this value convert to compressibility and integrate over the shaded gray area to get the thermodynamic gap size.

\subsubsection{Error Analysis of Compressibility and the Thermodynamic Gap}\label{sec:SI_Gap_Error}
 Each of the values used to determine the compressibility carries some uncertainty that must be propagated through to the final gap value. First, we derive the uncertainty of $\frac{\partial \mu}{\partial n}$. We define $K = \frac{\chi - \chi_{\textrm{band}}}{\chi_{\textrm{gap}}- \chi}$.

\begin{equation*}
    \sigma^2_{\frac{\partial \mu}{\partial n}} = e^2 \frac{1}{c_{\textrm{tg}} +c_{\textrm{bg}}} \left[\left(\frac{\partial K}{\partial \chi_{\textrm{band}}}\right)^2 \sigma^2_{\chi_{\textrm{band}}}     + \left(\frac{\partial K}{\partial \chi_{\textrm{gap}}}\right)^2 \sigma^2_{\chi_{\textrm{gap}}} + \left(\frac{\partial K}{\partial \chi_{p}}\right)^2 \sigma^2_{\chi_{p}}\right]
\end{equation*}

Now, to calculate the error in the thermodynamic gap, we must take into account the sum shown above. To do this we use a procedure similar to that shown in previous papers \cite{aronson_displacement_2025}. It is useful now to rewrite $m_t$ and $m_b$ as, 

$$m_t = \frac{r(\chi_{\textrm{band}} - \chi)}{\chi + r\chi_{\textrm{band}} - (1+r)\chi_{\textrm{gap}}}$$

$$m_b = \frac{\chi_{\textrm{band}} - \chi}{r\chi + \chi_{\textrm{band}} - (1+r)\chi_{\textrm{gap}}}$$
Here we have substituted in $c_q^{-1} = \frac{\chi - \chi_{\textrm{band}}}{\chi_{\textrm{gap}}- \chi} \frac{1}{c_{\textrm{tg}} + c_{\textrm{bg}}}$ and $r\equiv \frac{c_{\textrm{tg}}}{c_{\textrm{bg}}}$. Writing $m_t$ and $m_b$ in this manner shows that the gate capacitances enter only through the ratio $r$. Since we are dealing with a discrete data set we convert the integrals to discrete sums over the data. The variance of the thermodynamic gap can be written as, 

\begin{equation*}
\begin{split}
    \sigma_{\Delta\mu}^2 &= \left(\sum_i \frac{\partial m_{\textrm{t}}^{(i)}}{\partial \chi_{\textrm{band}}} \Delta V_{\textrm{tg}}^{(i)} + \sum_i \frac{\partial m_{\textrm{b}}^{(i)}}{\partial \chi_{\textrm{band}}} \Delta V_{\textrm{bg}}^{(i)}\right)^2 \sigma_{\chi_{\textrm{band}}}^2 + \left(\sum_i \frac{\partial m_{\textrm{t}}^{(i)}}{\partial \chi_{\textrm{gap}}} \Delta V_{\textrm{tg}}^{(i)} + \sum_i \frac{\partial m_{\textrm{b}}^{(i)}}{\partial \chi_{\textrm{gap}}} \Delta V_{\textrm{bg}}^{(i)}\right)^2 \sigma_{\chi_{\textrm{gap}}}^2 \\
    &+\left(\sum_i \frac{\partial m_{\textrm{t}}^{(i)}}{\partial r} \Delta V_{\textrm{tg}}^{(i)} + \sum_i \frac{\partial m_{\textrm{b}}^{(i)}}{\partial r} \Delta V_{\textrm{bg}}^{(i)}\right)^2 {\sigma_{r}}^2 \\
    &+ \sum_k \sum_l \left(\frac{\partial m_{\textrm{t}}^{(k)}}{\partial \chi} \Delta V_{\textrm{tg}}^{(k)} + \frac{\partial m_{\textrm{b}}^{(k)}}{\partial \chi} \Delta V_{\textrm{bg}}^{(k)}\right) \times \left(\frac{\partial m_{\textrm{t}}^{(l)}}{\partial \chi} \Delta V_{\textrm{tg}}^{(l)} + \frac{\partial m_{\textrm{b}}^{(l)}}{\partial \chi} \Delta V_{\textrm{bg}}^{(l)}\right)\textrm{Cov}(\chi^{k}, \chi^{l})
\end{split}
\end{equation*}

The first two terms account for the error in $\chi_{\textrm{gap}}$ and $\chi_{\textrm{band}}$. The third term accounts for error in the ratio of the top and bottom gate capacitances. The fourth term accounts for the point to point variation in the measured capacitance ratio $\chi$. Here we assume that the error in $\chi$ is uncorrelated such that we ignore all terms except for  $k = l$. Then our final term simply collapses to a single sum that looks like, 

$$\sum_k \left(\frac{\partial m_{\textrm{t}}^{(k)}}{\partial \chi} \Delta V_{\textrm{tg}}^{(k)} + \frac{\partial m_{\textrm{b}}^{(k)}}{\partial \chi} \Delta V_{\textrm{bg}}^{(k)}\right)^2 \sigma_{\chi}^2$$

Here we estimate the uncertainty of $\chi_{\textrm{gap}}$, $\sigma_{\chi_{\textrm{gap}}}$, using the mean squared error of the measured capacitance ratio as we gate across the band gap that exists at high displacement field and zero density. Since we are taking $\chi_{\textrm{band}}$ to be the flat region around the gap, as shown in Fig \ref{SI_GapExample}, we take the standard deviation over this region as $\sigma_{\chi_{\textrm{band}}}$. We take ${\sigma_r}$ to be $2\%$ of the measured ratio to account any uncertainty due to the finite thickness of the graphene. $\sigma^2_\chi$ is the variance in $\chi_{\textrm{band}}$ which tells us the point to point variability of the capacitance ratio in the region which we are measuring. We note that some of the gaps such as $\nu = 2/3$ as shown in Fig. \ref{SI_GapExample} sit on a slightly non-uniform background. Rather than trying to fit a line or polynomial to this background, we take the average between the two sides as shown in \ref{SI_GapExample}. This increases the uncertainty in $\chi_{\textrm{band}}$ and thus the uncertainty in the gap size. 

\subsubsection{$\nu =1$ Gap Extraction}\label{sec:SI_v1_gap}
\begin{figure}
    \centering
    \includegraphics[width=0.5\linewidth]{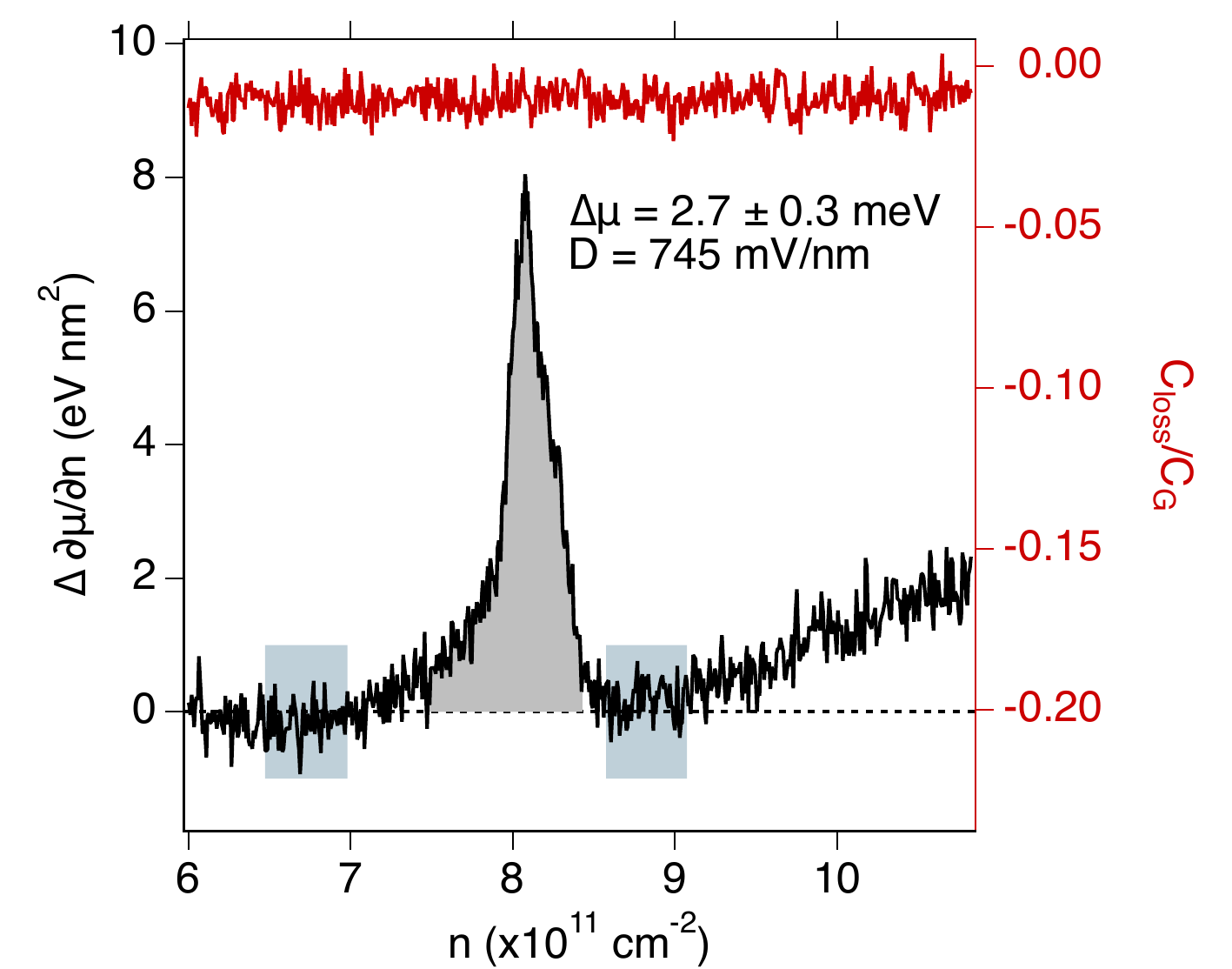}
    \caption{Compressibility and $C_{\textrm{loss}}$ at $D = \SI{745}{\milli \volt \per \nano \meter}$ at $T = \SI{320}{\milli \kelvin}$ and $f = \SI{38}{\kilo \hertz}$. We do not see any $C_{\textrm{loss}}$ that accompanies the peak in the inverse density of states. This ensures that the }
    \label{v1_GapExample}
\end{figure}

Figure \ref{ExtendedData_v1_Gap}b shows the thermodynamic gap extracted for filling factor $\nu =1$ at $T = \SI{320}{\milli \kelvin}$. Due to the larger gap, we measure this gap at a lower frequency of $\SI{38}{\kilo \hertz}$ in order to prevent enhancement of the penetration signal due to a large in-plane-resistance. An example of the gap extraction is shown in \ref{v1_GapExample}.

As discussed in for the fractional states, the $\nu =1$ gap resides on a non-uniform background of compressibility. As with the fractional case, we take the average between the two sides. This increases the uncertainty in the gap size.

\subsection{Capacitance Normalization}\label{sec:SI_Normalization}
In order to convert to thermodynamic compressibility, the frequency must be low enough so that the sample can charge and discharge on an AC time scale. If the frequency is too high, the sample cannot full charge. The penetration capacitance is no longer a probe of the density of states but instead a frequency dependent measurement of the compressibility and in-plane resistance. In this limit, large penetration signals a resistive state, and large $C_P$ can no longer be simply interpreted as in increase in the inverse compressibility. In this high-frequency limit, the contribution of the in-plane resistance will also give rise to a dissipative signal, $C_{\textrm{loss}}$, that is \qty{90}{\degree} out of phase with the excitation. In situations where large areas of the map have a substantial dissipative component, rather than converting the capacitance ratio to compressibility, we report the ratio of the penetration capacitance (the in-phase component) to the sample geometric capacitance (the penetration capacitance when the sample is in a band gap and cannot charge): $$C_P/C_G = \frac{\chi - \chi_{\textrm{band}}}{\chi_{\textrm{gap}}-\chi_{\textrm{band}}}$$
Thus, this normalized capacitance ratio is the penetration capacitance measured divided by the full geometric capacitance of the device. In other words, this can be considered to be the percentage of the electric field that penetrates the device.

\subsection{Frequency Dependence}\label{sec:SI_frequencyDependence}
\begin{figure*}
    \centering
    \includegraphics[width = 0.9 \linewidth]{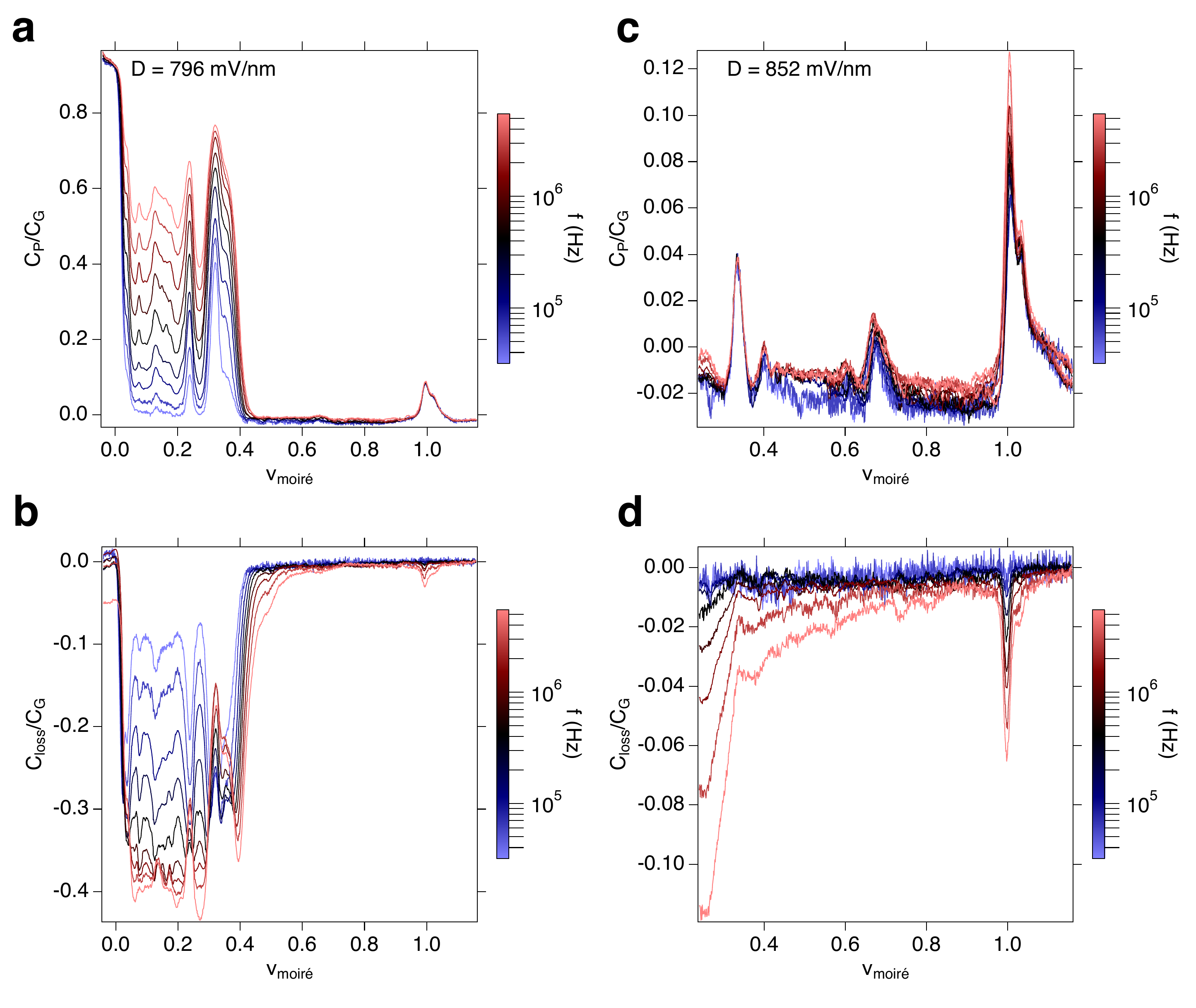}
    \caption{\justifying{Frequency dependence of the capacitance measurements at $T = \SI{320}{\milli \kelvin}$. \textbf{a} shows the in-phase or capacitive signal at $D = \SI{852}{\milli \volt \per \nano \meter}$ and \textbf{b} shows the dissipative signal or $C_{\textrm{loss}}$.  \textbf{c} shows the penetration capacitance at a lower displacement field of $D = \SI{796}{\milli \volt \per \nano \meter}$. \textbf{d} shows the dissipative signal or $C_{\textrm{loss}}$. We see that even at the lowest measured frequencies we observe a significant dissipative signal associated with each of the CDW states.}}
    \label{SI_fWaterfall}

\end{figure*}
When performing the capacitance measurements we measure both the in-phase signal, $C_P$, and also the out-of-phase signal, $C_{\textrm{loss}}$. As discussed in the main text, in the low-frequency regime the sample can charge and discharge within the period of the excitation, and the penetration capacitance can be expressed as $c_P^{-1} = c_{\textrm{tg}}^{-1} + c_{\textrm{bg}}^{-1} + \frac{\partial n}{\partial \mu} \frac{e^2}{c_{\textrm{tg}} c_{\textrm{bg}}}$. To first order, this expression is true when the the resistance of the sample is low enough such that $f \lesssim 1/(R C_G)$. However, as the frequency or the sample resistance increases, the low frequency limit is no longer valid. In this effective high-frequency regime, the out out-of-phase or dissipative component, $C_{\textrm{loss}}$, also begins to appear. When $C_{\textrm{loss}}$ becomes non-zero, the penetration capacitance no longer follows a simple relationship with the density of states. Instead, the penetration capacitance is dependent on both the density of states as well as the in-plane resistance. 

To illustrate this point, we measure the penetration capacitance at many different frequencies as shown in Fig. \ref{ExtendedData_nvdDvdf_maps}. In these maps the penetration capacitance that we measure in the WC region increases as the frequency increases, as expected for states with a large resistance. However, the FQAH states at higher displacement field have lower in-plane resistance and $C_P$ does not increase in magnitude as a function of frequency over our experimental range up to $\SI{10}{\mega \hertz}$. To further illustrate this point, we plot a line cut of the capacitance and loss at $D = \SI{852}{\milli \volt \per \nano \meter}$ where we observe FQAH states in Fig. \ref{SI_fWaterfall} \textbf{c} and \textbf{d}. Note, the $\nu =1$ gap displays a dissipative component that increases with frequency, as we might expect due to the larger gap of $\nu = 1$. Figure \ref{SI_fWaterfall} \textbf{a} and \textbf{b} show the penetration capacitance and loss at a lower displacement field of $D = \SI{796}{\milli \volt \per \nano \meter}$, where we observe trivial CDW states. The penetration capacitance displays a strong frequency dependence and grows with increasing frequency as expected for highly-resistive states. While the CDW states have a thermodynamic gap, the large dissipative component prevents proper gap extraction.

\begin{figure}
    \centering
    \includegraphics[width=0.95\linewidth]{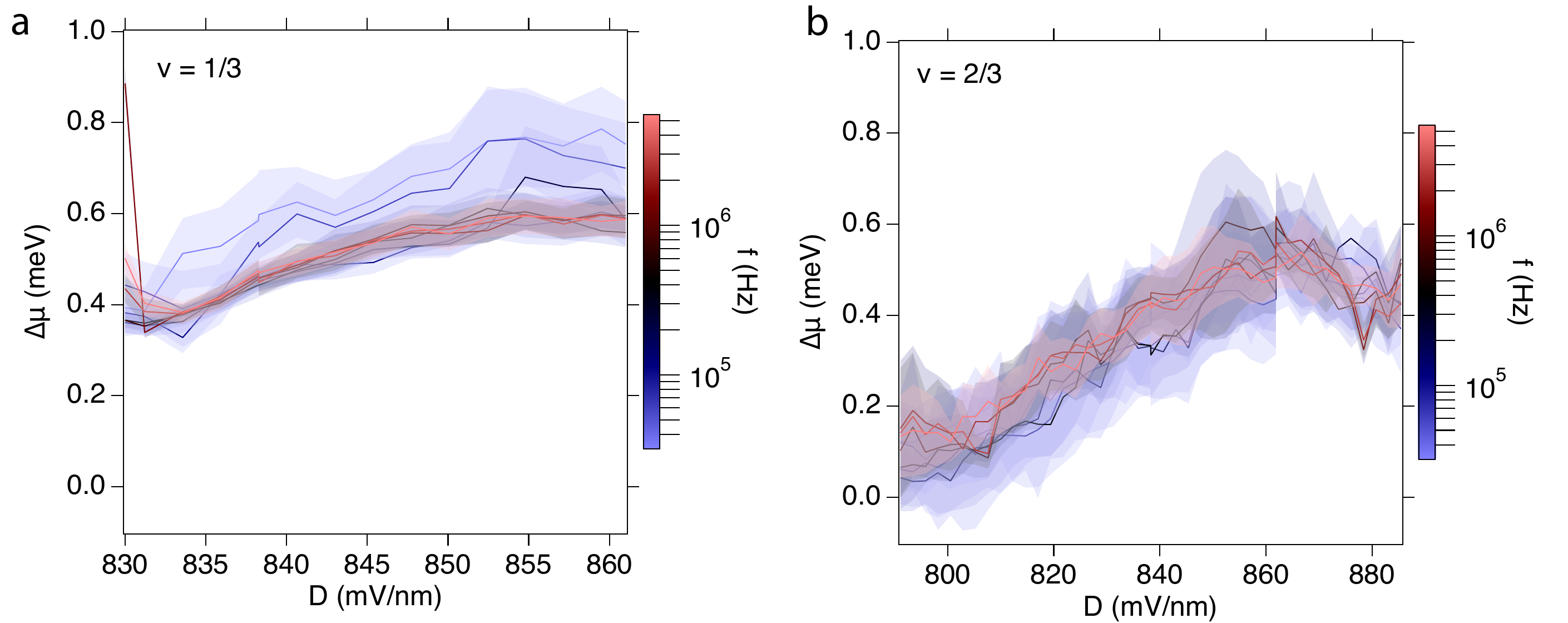}
    \caption{\justifying{Gap Extraction as a function of Frequency \textbf{a} Thermodynamic gap of $\nu = 1/3$ FQAH state at $\SI{320}{\milli \kelvin}$. \textbf{b} Thermodynamic gap of $\nu = 2/3$ FQAH state at $\SI{320}{\milli \kelvin}$. }}
    \label{gap_vs_f}
\end{figure}

To further verify that our choice of frequency does not effect the value of the gap size that we extract, we perform gap extraction on the series of maps as shown in Fig. \ref{ExtendedData_nvdDvdf_maps} for both $\nu = 1/3$ and $\nu = 2/3$. As shown in Fig. \ref{gap_vs_f}, the gap size of both $\nu = 1/3$ and $\nu = 2/3$ FQAH states does not increase as a function of frequency. If the frequency were too large so that the in-plane resistance were increasing $C_P$, one would expect the gap to increase as a function of frequency. Finally, we observe negligibly small $C_{\textrm{loss}}$ signals for all of the FQAH states.

In Fig. \ref{PhaseDiagram}\textbf{a} and \ref{FCI_LandauFan}\textbf{a} we plot the penetration capacitance measured at $f = \SI{10}{\mega \hertz}$. We observe that at $\SI{320}{\milli \kelvin}$, background features that we attribute to a bad contact, are minimized by measuring at higher frequencies. Figure \ref{ExtendedData_nvdDvdf_maps} shows the frequency dependence of the background features and their gradual diminishment with increased frequency. One explanation for this observation is that at higher frequencies the $R C$ time for charging the sample through the bad contact is significantly larger than the AC period. Hence, the bad contact stops participating in charging and discharging of the sample, and extraneous background features are minimized. Another possibility is that, at higher frequencies, the admittance of the bad contact is no longer dominated by its resistance, as it now also has a larger admittance contribution from its contact capacitance to the flake, also resulting in diminished features from the bad contact. Thus, at higher frequencies we see fewer features that we attribute to the bad contact compared to lower frequencies where we see features that partially obscure the the $2/5$ and $3/5$ FQAH states. In contrast, at dilution refrigerator temperatures, the resistance of the bad contact grows so large that we can measure at $f = \SI{202}{\kilo \hertz}$ with minimal obscuration of the FQAH states from background features generated by the bad contact.

\begin{figure}
    \centering
    \includegraphics[width=0.9\linewidth]{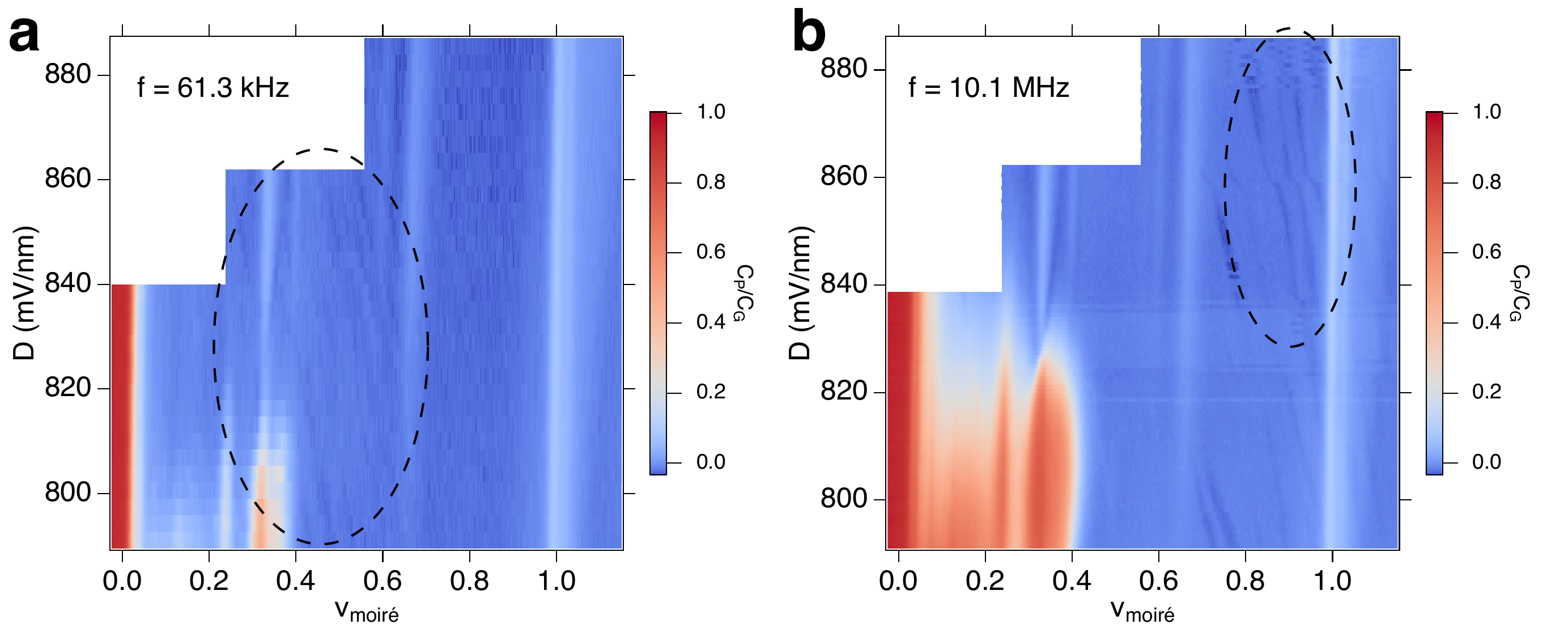}
    \caption{\justifying{\textbf{a)} $n-D$ phase diagram measures at $\SI{61.3}{\kilo \hertz}$ and $T = \SI{320}{\milli \kelvin}$ as also plotted in Fig. \ref{ExtendedData_nvdDvdf_maps}. The circled region shows enhanced diagonal features due to the poor contact. \textbf{b)} $n-D$ phase diagram measured at \SI{10.1}{\mega \hertz} as also plotted in Fig. \ref{PhaseDiagram}. The circled region denotes contact features that are also present at high frequencies but have moved to slightly higher densities and have a diminished impact.}}
    \label{ContactFeatureIllustration}
\end{figure}

\subsection{Excitation Dependence}\label{sec:SI_ExcitationDependence}
\begin{figure*}[ht]
    \centering
    \includegraphics[width= 0.7\linewidth]{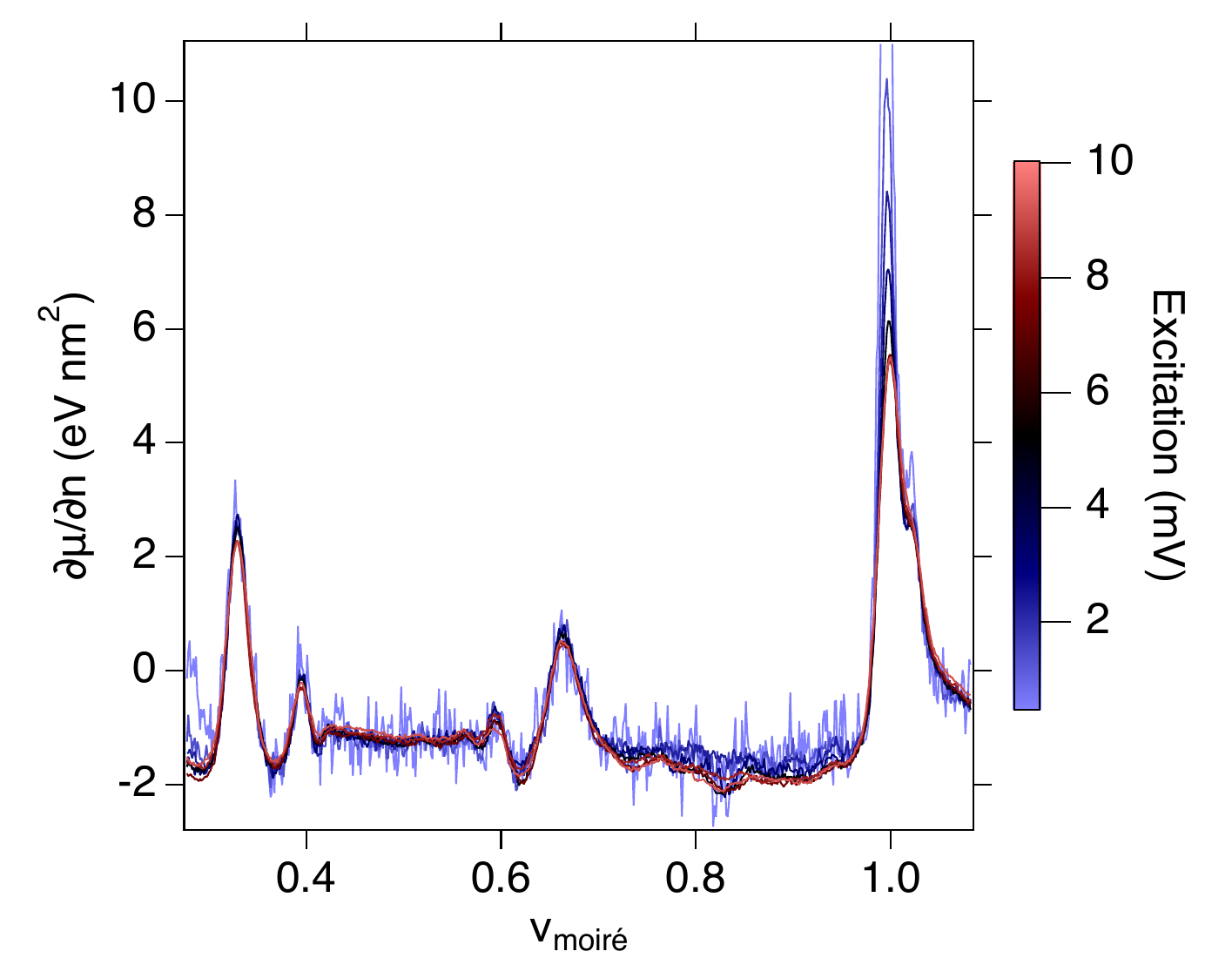}
    \caption{\justifying{Compressibility line cut at $ D = \SI{855}{\milli \volt \per \nano \meter}$ at $T = \SI{48}{\milli \kelvin}$ and $f = \SI{202}{\kilo \hertz}$ as a function of excitation size. Increasing excitation size does not significantly broaden the fractional states. However, it does decrease the height of the peak at $\nu =1$ at these very low temperatures.}}
    \label{SI_excDependence}
\end{figure*}

In order to ensure that we use a sufficiently small excitation amplitude, we take traces at a series of different amplitudes as shown in Fig. \ref{SI_excDependence}. Up to $V_{\textrm{rms}} = \SI{10}{\milli \volt}$, we do not see observable broadening of the peaks in the inverse compressibility associated with the FQAH states.All plots at $T = \SI{39}{\milli \kelvin}$ use en excitation amplitude of $V_{\textrm{rms}} = \SI{7}{\milli \volt}$. However, we do see that the peak in $\frac{\partial \mu}{\partial n}$ at $\nu = 1$ becomes diminished even at the smallest amplitudes. Thus, we do not quantitatively extract the jump in chemical potential at $\nu = 1$ at $\SI{39}{\milli \kelvin}$. Additionally, $\nu =1$ develops a significant dissipative signal preventing proper gap extraction at $\SI{202}{\kilo \hertz}$.

\section{Determination of Chern Numbers from the Landau Fan}\label{sec:SI_Chern_Number_Extraction}

\begin{table*}
    \centering
    \renewcommand{\arraystretch}{1.5}
    \begin{tabular}{|c|c|c|c|}
        \hline
        $\nu$ (Magnetic Field Fit Range)  & Gaussian + Linear Peak Finding &  Gaussian Peak Finding &  Maximum Peak Finding   \\
        \hline
        $1/3$ (\SIrange{0}{4}{\tesla}) & $0.329\pm 0.001$&$0.335\pm 0.002$ & $0.35  \pm 0.01$ \\
        
        $2/5 \textrm{ }(\SIrange{0}{1.75}{\tesla})$ & $0.465 \pm 0.007$ & $0.477 \pm 0.006$ &   $0.48 \pm 0.04$ \\

        $3/5 \textrm{ }(\SIrange{0}{3}{\tesla})$ &  $0.602 \pm 0.007$& $ 0.640 \pm 0.007$ & $0.63 \pm 0.03$ \\

        $2/3 \textrm{ } (\SIrange{0}{4}{\tesla})$ &  $0.691 \pm 0.002$&  $0.691 \pm  0.002$ & $0.69 \pm 0.02$ \\
        
        $1 \textrm{ }(\SIrange{0}{4}{\tesla})$ &  $1.017 \pm 0.002$& $1.033 \pm 0.002$&   $1.005  \pm  0.006$\\
        \hline
    \end{tabular}
    \caption{Calculated slopes for the fan shown in Fig \ref{FCI_LandauFan} using three different methods of peak finding. These are statistical error bars that report the $95 \%$ confidence interval oly. They do not include systematic effects as described in the text of the supplement.}
    \label{tab:scan207_Slopes}
\end{table*}

We determine determine the Chern number of each of the gapped fractional states by  first determining the peak position of each of the fractions at each magnetic field. We do this in three different ways. The first method is simply by determining the location of the maximum penetration capacitance. The second is by fitting over a window in density with a gaussian to find the center. Lastly, we use a gaussian with an added linear background to find the center. After obtaining the location at each field, we perform a linear fit to the location in density as a function of magnetic field to obtain the slope $\partial n/\partial B$. The results of each of the fitting methods is very similar and shown in Table \ref{tab:scan207_Slopes}.

For the both the gaussian fit and the gaussian fit plus a linear background, we take the standard deviation of the center and use it to weight our linear fits of the Landau fan. For the fit to the maximum, the error is simply the error in the error in the slope parameter of the linear fit. The error bars as shown in Table \ref{tab:scan207_Slopes} are statistical error bars only. They do not take into account systematic error such as error in the capacitance per unit area ($c_{\textrm{tg}}$ or $c_{\textrm{bg}}$), or any corrections due to a finite density of states.

\section{Determination of the Moiré Wavelength and Filling Factor}
To determine the twist angle we fit both the $\nu =1/3$ and $\nu = 1$ peaks in capacitance. The difference in density, $\Delta n$, can then be used to calculate the moire wavelength, $\lambda_{\textrm {moiré}} = \sqrt{\frac{2}{3 \times (3/2 \Delta n)}}$. To convert from density to filling factor, $\nu_{\textrm{moiré}}$, at high displacement field we use, $\nu_{\textrm{moiré}} = (n-n^*_{\textrm{off}})/n_1$, where $n_1 = 3\times \Delta n$ and $n^*_{\textrm{off}}$ is the density offset that is caused by the band gap expanding as a function of displacement field. $n^*_{\textrm{off}}$ is chosen to align the incompressible state with $\nu =1$.

\section{Transport Measurements } \label{sec:SI_Transport}
    \subsection{Transport Measurements for D1}
    \begin{figure}
        \centering
        \includegraphics[width=0.5\linewidth]{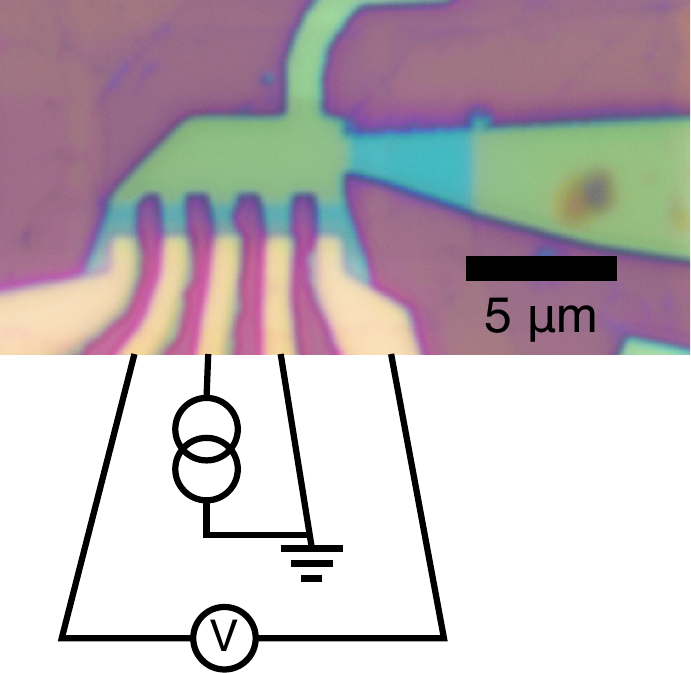}
        \caption{\justifying{Picture of device and configuration of the four-terminal transport measurement on sample D1. The drain contact here is kept grounded in the refrigerator.}}
        \label{fig:placeholder}
    \end{figure}
See transport details in the methods section.

\putbib
\end{bibunit}
\newpage
\end{document}